\documentclass[aps,prd,amsmath,twocolumn,showpacs,nofootinbib]{revtex4-1}

\usepackage{graphicx}
\usepackage{hyperref}
\usepackage{color}

\hypersetup{colorlinks=true}

\let\l=\left
\let\r=\right
\newcommand{\be}{\begin{equation}}
\newcommand{\ee}{\end{equation}}
\newcommand{\bea}{\begin{eqnarray}}
\newcommand{\eea}{\end{eqnarray}}
\newcommand{\D}{\mathrm{d}}  
\newcommand{\lB}{-\!\!\!\!\!\lambda_{\rm b}}

\begin{document}

\title{Models of rotating boson stars and geodesics around them: new type of orbits}

\author{Philippe Grandcl\'ement}
\email{philippe.grandclement@obspm.fr}
\affiliation{
Laboratoire Univers et Th\'eories, UMR 8102 du CNRS,
Observatoire de Paris, Universit\'e Paris Diderot, F-92190 Meudon, France}
\author{Claire Som\'e}
\email{claire.some@obspm.fr}
\affiliation{
Laboratoire Univers et Th\'eories, UMR 8102 du CNRS,
Observatoire de Paris, Universit\'e Paris Diderot, F-92190 Meudon, France}
\author{Eric Gourgoulhon}
\email{eric.gourgoulhon@obspm.fr}
\affiliation{
Laboratoire Univers et Th\'eories, UMR 8102 du CNRS,
Observatoire de Paris, Universit\'e Paris Diderot, F-92190 Meudon, France}

\date{4 July 2014}  

\begin{abstract} 
We have developed a highly accurate numerical code capable of solving 
the coupled Einstein-Klein-Gordon system, in order to construct rotating boson 
stars in general relativity. Free fields and self-interacting fields,
with quartic and sextic potentials, are considered. In particular, 
we present the first numerical 
solutions of rotating boson stars with rotational quantum number $k=3$ and 
$k=4$, as well as the first determination of the maximum mass of free-field boson
stars with $k=2$. 
We have also investigated timelike geodesics in the spacetime generated
by a rotating boson star for $k=1$, $2$ and $3$. A numerical integration of the 
geodesic equation has enabled us to identify a peculiar type of orbits: the
zero-angular-momentum ones. These orbits pass very close to the center 
and are qualitatively different from orbits around a Kerr black hole. 
Should such orbits be observed, they would put stringent constraints on 
astrophysical compact objects like the Galactic center.
\end{abstract} 


\maketitle


\section{Introduction} \label{s:intro}

Boson stars are localized configurations of a self-gravitating complex 
scalar field, introduced in the end of the sixties by Bonazzola and Pacini \cite{BonazP66}, 
Kaup \cite{Kaup68} and Ruffini and Bonazzola \cite{RuffiB69}. 
Motivated by the facts that (i) 
boson stars are, at the fundamental level, the simplest self-gravitating 
configurations of ``matter''
and (ii) they can act as black hole mimickers \cite{GuzmaR09}, 
numerous studies of boson stars have been performed
(see \cite{Jetze92,LeeP92,SchunM03,LieblP12} for a review). 
A recent impetus to the topic has been provided by the discovery 
of the Higgs boson at CERN \cite{ATLAS-CMS12}, which proves the
existence of fundamental scalar fields in Nature. 
In addition, the main paradigm of current primordial cosmology, the inflation, is generally 
based on a scalar field (the 
\emph{inflaton}) \cite{PeterU12}. 
Still in the field of cosmology, we note that many dark energy models also 
rely on a scalar field, such as the quintessence model \cite{Tsuji13}.

Boson stars studies have explored a large parameter space \cite{Jetze92,LeeP92,SchunM03,LieblP12}, 
by varying
the scalar field's self-interaction potential, the 
spacetime symmetry (static, axisymmetric rotating or  
dynamical configurations), the number of
spacetime dimensions (2 to 5), the spacetime asymptotic (flat or AdS) 
or the theory 
of gravity (general relativity, Einstein-Gauss-Bonnet gravity, scalar-tensor gravity, etc.). 

We consider boson star models with a minimal coupling of the scalar field to gravity. They are described by the following action \footnote{We use geometrized units, 
in which both Newton gravitational constant $G$ and the speed of light $c$ are set to unity:
$G=c=1$. We also use the convention $(-,+,+,+)$ for the spacetime metric signature.}:
\be\label{e:action}
S = \int  \left (\mathcal{L}_g  + \mathcal{L}_\Phi \right) \sqrt{-g} \, \D^4 x,
\ee
where 
$\mathcal{L}_g$ is the Hilbert-Einstein Lagrangian of the gravitational field:
\be
{\mathcal L}_g = \frac{1}{16\pi} R, 
\ee
$R$ being the Ricci scalar associated with the spacetime metric $g_{\alpha\beta}$,
and $\mathcal{L}_\Phi$ is the Lagrangian of the complex scalar field $\Phi$:
\be \label{e:scalar_Lag}
{\mathcal L}_\Phi = -\frac{1}{2} \l[\nabla_\mu \Phi \nabla^\mu \bar{\Phi} + V\l(\l|\Phi\r|^2\r) \r],
\ee
The potential $V$ is assumed to depend on $\l|\Phi\r|^2$ only (U(1) symmetry). 
The simplest choice for $V$ is that 
corresponding to a \emph{free field}: 
\be \label{e:V_free_field}
    V(|\Phi|^2) = \frac{m^2}{\hbar^2} |\Phi|^2 ,  
\ee
the constant $m$ being the boson mass. 
Boson stars built on (\ref{e:V_free_field}) are called \emph{mini-boson stars}
\cite{SchunM03}, 
for their maximum mass is very small, except for extremely tiny values of $m$
\cite{Jetze92}.
To get massive boson stars, Colpi et al. \cite{ColpiSW86} have added 
a repulsive self-interacting term of the type $\Lambda |\Phi|^4$ to the potential $V$:
\be
\label{e:V_phi_4}
V(|\Phi|^2) = \frac{m^2}{\hbar^2} |\Phi|^2 \left( 1 + 2\pi \Lambda |\Phi|^2
    \right) , 
\ee
where $\Lambda$ is a positive constant. Important mass can then be reached in the limit $\Lambda \gg 1$.
Another generalization of the potential has been proposed by Friedberg et al. \cite{FriedLP87}:
\be
\label{e:V_sol}
V(|\Phi|^2) = \frac{m^2}{\hbar^2} |\Phi|^2 \left(1 - \frac{|\Phi|^2}{\sigma^2}\right)^2,
\ee
where $\sigma$ is a constant, which corresponds to the value of the field in 
the degenerate vacuum state. The potential (\ref{e:V_sol}) authorizes
localized configurations of \emph{solitonic} type, i.e. 
that can exist even in the absence of gravity, which is not possible for a
free field. 
A related alternative is based on a potential of the form \cite{VolkoW02}:
\be
\label{e:V_Q-balls}
V\left(|\Phi|^2\right) = \frac{m^2}{\hbar^2} |\Phi|^2 + \lambda \left(|\Phi|^6 - a |\Phi|^4 \right), 
\ee 
where $\lambda$ and $a$ are two constants.
The corresponding solutions in flat spacetime are called \emph{$Q$-balls} 
(see e.g. \cite{TamakS10,TamakS11}). 

Beside the choice of the potential $V$, 
models of stationary and axisymmetric rotating boson stars are based on the following ansatz for the 
complex scalar field $\Phi$:
\be
\label{e:ansatzaxe}
\Phi = \phi \l(r,\theta\r) \exp\l[i\l(\omega t - k\varphi\r)\r],
\ee
where $(t,r,\theta,\varphi)$ are coordinates adapted to the spacetime symmetries
(i.e. $\partial/\partial t$ is the stationarity generator  and 
$\partial/\partial\varphi$ the axisymmetry one), 
$\phi = | \Phi|$ is a positive real function of $r$ and $\theta$ only, 
$\omega$ is a real constant
and $k$ is an integer, called the \emph{rotational quantum number} \cite{KleihKL05,LieblP12}. 
Note that $k$ has to be an integer in order for the scalar field to be single-valued
at $\varphi=0$ and $\varphi=2\pi$. 
The ansatz (\ref{e:ansatzaxe}) has been introduced by Kaup \cite{Kaup68} 
for $k=0$ (nonrotating case) and by Schunck and Mielke \cite{SchunM96}
for $|k|\geq 1$. It leads to stationary solutions for the spacetime metric. 

The first (numerical) solutions for rotating boson stars in general relativity
have been obtained by Schunck and Mielke in 1996 \cite{SchunM96,SchunM98}, 
considering a free scalar field [i.e. $V$ given by Eq.~(\ref{e:V_free_field})] 
and $k=1$ to $10$, as well as 
$k=500$, in Eq.~(\ref{e:ansatzaxe}); 
their study was limited to the weakly relativistic regime. 
The strongly relativistic regime has been tackled in the works of 
Ryan \cite{Ryan97} ($|\Phi|^4$ self-interaction and approximation valid 
for $k\gg 1$)
and Yoshida and Eriguchi \cite{YoshiE97b} (free field with $k=1$ and $k=2$).
In particular, the latter authors have performed the first determination of
the maximum mass of free-field rotating boson stars for $k=1$. 
In 2004, Lai computed a full sequence for $k=2$, thereby 
obtaining a maximum mass value, but his code suffered from regularity issues
at the rotation axis. In particular the maximum mass for $k=1$ was significantly 
different from that obtained by Yoshida and Eriguchi and Lai's 
results have not been published but in the PhD thesis \cite{Lai04}. 
In the present work, we confirm the value found by Yoshida and Eriguchi 
[Eq.~(\ref{e:M_max_k_1}) below]. 
In 2005, Kleihaus et al. \cite{KleihKL05}  have computed 
rotating boson stars with the self-interacting potential (\ref{e:V_Q-balls}) 
for $k=1$, generalizing the rotating $Q$-balls models of Volkov and W\"ohner 
\cite{VolkoW02} to the self-gravitating case. 
They extended the study to $k=2$ and to negative parity scalar fields
(i.e. $\Phi$ antisymmetric with respect to the equatorial plane $\theta=\pi/2$)
 in \cite{KleihKLS08} and analyzed the stability of the configurations 
 in \cite{KleihKS12}.
For the sake of completeness, let us mention that Hartmann et al. have
studied special cases of rotating boson stars in 5-dimensional spacetimes
\cite{HartmKKL10} (the boson field is then actually a doublet of complex
scalar fields) by assuming that the two angular momenta (associated with 
the 2 independent planes of rotations in 5 dimensions) are equal. 
Their results have been extended recently to 5-dimensional Einstein-Gauss-Bonnet
gravity \cite{BrihaR13}. Solutions in higher dimensions, with only one Killing vector,
 are obtained in \cite{StotyLOHM14, HendeMS14}.
Recently Herdeiro and Radu \cite{HerdeR14} constructed rotating solutions containing, 
in addition to the scalar field, an event horizon. Their solutions are 
thus hairy black holes, which can be viewed as intermediate states between rotating boson stars and Kerr black holes.

In the present article, we have considered both free-field boson stars
[potential (\ref{e:V_free_field})] and self-interacting-field ones, 
based on the potentials (\ref{e:V_phi_4}) and (\ref{e:V_sol}), 
with the rotational quantum number ranging from
$k=0$ to $k=4$. We have developed a new numerical code, based on a spectral 
method, to compute the solutions of the coupled Einstein-Klein-Gordon 
equations. We have also investigated the timelike geodesics in the obtained
numerical spacetimes. To our knowledge, the determination of geodesics 
around a rotating boson star has never been performed before; only the
case of geodesics around nonrotating spherically symmetric boson stars has been 
dealt recently by Diemer et al. \cite{DiemeEHST13}, for a self-interacting potential which 
reduces to (\ref{e:V_Q-balls}) in the weak field limit. The particular case of 
circular timelike geodesics around static boson stars
with various types of self-interaction (free, $|\Phi|^4$, and solitonic) has been 
investigated also recently by Macedo et al. \cite{MacedPCC13}.

This article is organized as follows.
In Sec.~\ref{s:field_equations}, we present the field equations to be solved (Einstein-Klein-Gordon system)
as well as the relevant global quantities. 
Section~\ref{s:spheric} focuses on models of 
nonrotating spherically symmetric
boson stars, 
while Sec.~\ref{s:rot} presents the models with rotation, which are axisymmetric. 
In both cases a detailed description  of the numerical method, based on spectral methods, is given. Various error indicators are also exhibited and discussed. 
Section~\ref{s:orbits} is devoted to the study of orbits of massive particles around boson stars. 
Circular orbits and zero-angular momentum one are discussed, the latter ones being computed via
a numerical integration of the geodesic equation. 
Conclusions and perspectives are given in Sec.~\ref{s:ccl}.


\section{Field equations and global quantities} \label{s:field_equations}

\subsection{Equations to be solved}

Variation of the action (\ref{e:action}) with respect to the spacetime metric $g_{\alpha\beta}$ leads to 
Einstein equation 
\be \label{e:Einstein_eq}
R_{\alpha\beta} -\frac{1}{2} R g_{\alpha\beta} = 8 \pi T_{\alpha\beta}, 
\ee 
where $R_{\alpha\beta}$ is the Ricci tensor associated with $g_{\alpha\beta}$, 
$R:=g^{\mu\nu} R_{\mu\nu}$ 
and $T_{\alpha\beta}$ is the energy-momentum tensor of the 
scalar field:
\be \label{e:tmunu}
T_{\alpha\beta} = \nabla_{(\alpha} \Phi \nabla_{\beta)} \bar{\Phi}
    - \frac{1}{2} \left[ \nabla_\mu \Phi \nabla^\mu \bar{\Phi} 
+ V \left(|\Phi|^2\right) \right] 
    g_{\alpha\beta} . 
\ee

Variation of the action (\ref{e:action}) with respect to the scalar field $\Phi$
results in the Klein-Gordon equation:
\be
\label{e:kg}
\nabla_\mu \nabla^\mu \Phi = \frac{{\rm d}V}{{\rm d}|\Phi|^2} \; \Phi.
\ee

Given some choice of the potential $V$, the field equations 
(\ref{e:Einstein_eq})-(\ref{e:kg}) are solved for $(g_{\alpha\beta},\Phi)$, 
under the assumptions of stationarity and axisymmetry for the spacetime
metric $g_{\alpha\beta}$ and the ansatz (\ref{e:ansatzaxe}) for $\Phi$
(which is compatible with the assumed spacetime symmetries). 

In the following, we use the language of the 3+1 formalism 
(see e.g. \cite{Gourg12,BaumgS10,Alcub08}), denoting by
$\Sigma_t$ the hypersurfaces of constant $t$, by $n^\alpha$ the timelike 
future-directed unit normal to $\Sigma_t$,  by $\gamma_{ij}$ the metric
induced by $g_{\alpha\beta}$ on $\Sigma_t$, by $N$ the lapse function 
and by $\beta^\alpha$ the shift vector, the last two quantities being 
defined by the orthogonal decomposition of the stationarity generator:
$\left( \partial/\partial t \right)^\alpha = N n^\alpha + \beta^\alpha$, with
$n_\mu \beta^\mu=0$. 
 The spacetime metric line element can then be written as
\be \label{e:metric_3p1}
  g_{\mu\nu} \, \D x^\mu \, \D x^\nu = 
  - N^2 \D t^2 + \gamma_{ij} (\D x^i + \beta^i \D t)
		(\D x^j + \beta^j \D t) . 
\ee 
Note that for stationary and axisymmetric spacetimes that are circular 
(cf. Sec.~\ref{s:rot_eq}), such as those of rotating boson stars with $\Phi$
of the type (\ref{e:ansatzaxe}), $\beta^i = (0,0,\beta^\varphi)$. 

\subsection{Global quantities} \label{s:global}

Via Noether's theorem, the U(1) symmetry of the Lagrangian (\ref{e:scalar_Lag}) 
yields the following conserved current \cite{Kaup68,RuffiB69,LieblP12}:
\be \label{e:def_j}
    j^\alpha = \frac{i}{2\hbar} \left( \bar\Phi \nabla^\alpha \Phi - \Phi \nabla^\alpha \bar\Phi
        \right) . 
\ee
It is divergence-free: $\nabla_\mu j^\mu = 0$ and its flux through a 
 hypersurface $\Sigma_t$ gives the \emph{scalar charge} or
\emph{total particle number} of the boson star:
\be \label{e:def_Q}
    \mathcal{N} := - \int_{\Sigma_t} n_\mu j^\mu \, \sqrt{\gamma} \, \D^3 x,
\ee
where 
$\gamma:=\det\gamma_{ij}$ (compare e.g. with Eq.~(4.4) of 
Ref.~\cite{Gourg10}). 
By plugging the ansatz (\ref{e:ansatzaxe}) into (\ref{e:def_j}), we get
\be
    j^\alpha = \hbar^{-1} \phi^2 \nabla^\alpha (k\varphi-\omega t) , 
\ee
so that (\ref{e:def_Q}) becomes
\be \label{e:Q_3p1}
    \mathcal{N} = \frac{1}{\hbar} 
        \int_{\Sigma_t} \frac{1}{N} \left(\omega + k \beta^\varphi
    \right) \phi^2 \, \sqrt{\gamma} \, \D^3 x . 
\ee

The spacetime symmetries lead to two other conserved quantities, expressed by
the Komar integral of the related Killing vector. 
The first one, associated to the Killing vector $\xi = \partial/\partial t$ 
is the \emph{gravitational mass} $M$ of the boson star. 
While the original Komar expression invokes a surface integral of the gradient of the
Killing vector, it can be rewritten as the following volume integral 
[see e.g. Eq.~(8.63) of Ref.~\cite{Gourg12}]:
\be
    M = 2 \int_{\Sigma_t} \left( T_{\mu\nu} n^\mu \xi^\nu
        - \frac{1}{2} T n_\mu \xi^\mu \right) \, \sqrt{\gamma} \, \D^3 x . 
\ee
In the present case, 
$T_{\mu\nu} n^\mu \xi^\nu = N^{-1} (T_{tt} - T_{t\varphi} \beta^\varphi)$.
Using the expressions derived in Appendix~\ref{s:app_T} for $T_{tt}$,
$T_{t\varphi}$ and $T$ [Eqs.~(\ref{e:T_tt}), (\ref{e:T_tp}) and 
(\ref{e:trace_T}) respectively], we arrive at
\be \label{e:grav_mass}
    M = \int_{\Sigma_t} \left[ \frac{2\omega}{N}(\omega + k\beta^\varphi)
    \phi^2 - N V \right] \, \sqrt{\gamma} \, \D^3 x . 
\ee

The second spacetime symmetry, the axisymmetry, leads to the
\emph{angular momentum} $J$. The Komar expression can
be recast as [see e.g. Eq.~(8.75) of Ref.~\cite{Gourg12}]
\be
    J = - \int_{\Sigma_t} T_{\mu\nu} n^\mu \chi^\nu
         \, \sqrt{\gamma} \, \D^3 x , 
\ee
where $\chi$ stands for the Killing vector $\partial/\partial\varphi$ 
Now $T_{\mu\nu} n^\mu \chi^\nu = N^{-1} (T_{t\varphi} - \beta^\varphi T_{\varphi\varphi})$. 
Using expressions (\ref{e:T_tp}) and (\ref{e:T_pp}) for $T_{t\varphi}$
and $T_{\varphi\varphi}$, we get
\be \label{e:J_vol_int}
    J = k \int_{\Sigma_t} \frac{1}{N} \left(\omega + k \beta^\varphi
    \right) \phi^2 \, \sqrt{\gamma} \, \D^3 x . 
\ee
Comparing with (\ref{e:Q_3p1}), we recover the quantization law
for the angular momentum of a rotating boson star \cite{SchunM96}:
\be \label{e:J_kQ}
    J = k \hbar \mathcal{N}. 
\ee

\subsection{Units and order of magnitude} \label{s:units}

As stated above, in this article we use geometrized units: $c=1$ and $G=1$. 
From Eqs.~(\ref{e:action})-(\ref{e:scalar_Lag}) and the fact that 
$R$ has dimension $\mathrm{length}^{-2}$, it is clear that
$\Phi$, and hence $\phi$, is dimensionless in these units. 
In non-geometrized units, the dimension of $\Phi$ is actually 
$\mathrm{mass}^{1/2}\times \mathrm{length}^{1/2}\times \mathrm{time}^{-1}$ (i.e. 
square-root of a force), 
so that $\tilde\Phi := (\sqrt{m}/\hbar)\,  \Phi$ has the dimension of a wave function, 
i.e. $\mathrm{length}^{-3/2}$. 

In view of (\ref{e:V_free_field}) or (\ref{e:V_sol}), a natural length
scale that appears in the problem is the \emph{boson reduced Compton 
wavelength}\footnote{In this section, we restore the $G$'s and $c$'s.}:
\be \label{e:def_lambda_Compton}
    \lB := \frac{\hbar}{m c} . 
\ee
The \emph{boson gravitational mass scale} associated with $\lB$
is
\be
    M_{\rm b} := \frac{c^2 \lB}{G} = \frac{m_{\rm P}^2}{m} , 
\ee
where $m_{\rm P} := \sqrt{\hbar c/G} \simeq 2.18\; 10^{-8} \; {\rm kg}$
is the Planck mass. Note that in geometrized units, 
$M_{\rm b} = \lB$. 

In the free scalar field case, the maximal mass allowed for a boson star is of order $M_{\rm b}$. As can be seen
in Tab. \ref{t:masses}, it leads to small masses, even if the scalar field has the same mass as the electron.

For the potential (\ref{e:V_phi_4}) the situation is different. Indeed one can show \cite{ColpiSW86} that the maximal mass scales as $M_{\rm max} \simeq \Lambda^{1/2} \displaystyle\frac{m_{\rm P}^2}{m} = \l(\displaystyle\frac{\lambda}{4\pi}\r) ^{1/2} \displaystyle\frac{m_{\rm P}^3}{m^2}$. $\lambda$ is the true coupling constant (see \cite{ColpiSW86}). If one assumes that $\lambda$ is close to one, it leads to a dramatic increase of the allowed masses as can be seen in Tab. \ref{t:masses}. In particular, if the scalar field has the same mass as the electron, one can reach values comparable to the ones of supermassive black holes. Let us point out that the corresponding values of $\Lambda$ are then very large, in particular much larger than the value studied in this paper (see Sec. \ref{s:nonfree} where $\Lambda=200$). Similar considerations hold for the potential (\ref{e:V_sol}) and we refer the reader to \cite{LeeP92} for more details.

\begin{table}
\begin{tabular}{l|c|c|c}
\hline
Scalar field mass & $m_{\rm Higgs}$ & $m_{\rm proton}$ & $m_{\rm electron}$ \\
\hline
Mass free-field (in kg) & $2 \cdot 10^9$ & $3\cdot 10^{11}$ & $5 \cdot 10^{14}$ \\
\hline
Mass potential with $\lambda \approx 1$ (in kg) & $2\cdot 10^{26}$  & $4\cdot 10^{30}$  & $1 \cdot 10^{37}$ \\
Value of $\Lambda$ & $7\cdot 10^{32}$ & $1\cdot 10^{37}$ & $5 \cdot 10^{43}$ \\
\hline
\end{tabular}  
\caption{ \label{t:masses}
Order of magnitude of the masses of various boson stars. The three columns correspond to various masses of the scalar field : the mass of the Higgs boson, the mass of the proton and the mass of the electron (we do not say the proton and the electron are bosons !). For the Higgs mass one uses  
$m = 125 \, {\rm GeV}$ \cite{ATLAS-CMS12}. The first line shows the values for the free field and the second one for the potential (\ref{e:V_phi_4}), assuming $\lambda \approx 1$. The last line gives the corresponding values of $\Lambda$. }
\end{table}


\section{Spherically symmetric models}\label{s:spheric}

\subsection{Equations}

Spherically symmetric solutions are constructed by setting $k=0$ in the ansatz (\ref{e:ansatzaxe})
and by demanding that the field modulus $\phi$ depends only on $r$: 
\be
\label{e:ansatz}
\Phi = \phi(r) \exp(i\omega t) . 
\ee
Note that according to Eq.~(\ref{e:J_kQ}), $k=0$ implies a vanishing angular momentum;  the
spherically symmetric solutions are thus nonrotating. 
The corresponding spacetime is static [while $\Phi$ is not, as it is clear
from (\ref{e:ansatz})]. In particular, $\beta^i=0$ in 
Eq.~(\ref{e:metric_3p1}). 
Note that, contrary to fluid stars, staticity does not imply 
that the boson stars have to be spherically symmetric, as demonstrated by 
the non-spherically symmetric static solutions obtained by 
Yoshida and Eriguchi \cite{YoshiE97a}. This is due to the anisotropy
of the scalar field energy-momentum tensor (\ref{e:tmunu}). Therefore the spherical
symmetry of our models results from the assumption $\phi = \phi(r)$ in 
(\ref{e:ansatz}). 

Thanks to spherical symmetry, 
we may choose spatial coordinates such that $\gamma_{ij} = \Psi^4 f_{ij}$
where $f_{ij}$ is a flat metric (\emph{isotropic coordinates}). 
The metric line element is thus
\be
  g_{\mu\nu} \, \D x^\mu \, \D x^\nu =  - N^2 \D t^2 + \Psi^4 
  \left[ \D r^2 + r^2 (\D\theta^2 + \sin^2\theta\, \D\varphi^2) \right] . 
\ee

The unknown functions are $N$, $\Psi$ and $\phi$, which all depend only 
on $r$ and obey the system obtained from the Einstein equation
(\ref{e:Einstein_eq}) and the Klein-Gordon equation (\ref{e:kg}):
\bea
& & \Delta_3 \Psi = - \pi \Psi^5 \l[ \l(\frac{\omega \phi}{N}\r)^2 + \frac{\partial \phi \partial \phi}{\Psi^4} + V\r] \label{e:eqpsisph} \\
& & \Delta_3 N + 2 \frac{\partial N \partial \Psi}{\Psi} = 4\pi N \Psi^4 \l(2\frac{\omega^2}{N^2} \phi^2 - V \r) 
    \label{e:eqnsph} \\
& &\Delta_3 \phi - \Psi^4 \l(\frac{{\rm d}V}{{\rm d}\l|\Phi\r|^2}-\frac{\omega^2}{N^2}\r) \phi =  - \frac{\partial \phi \partial N}{N} 
        - 2 \frac{\partial \phi \partial \Psi}{\Psi}, \nonumber \\
    & & \label{e:eqphisph}
\eea
where $\Delta_3 := \displaystyle\frac{\D^2}{\D r^2} + \frac{2}{r} \frac{\D}{\D r}$ (3-dimensional flat Laplacian in spherical symmetry)  and $\partial f \partial g := \displaystyle\frac{\D f}{\D r}\frac{\D g}{\D r}  $.

The system is closed by demanding that, at spatial infinity, one recovers Minkowski spacetime. 
This implies that $N\rightarrow 1$, $\Psi \rightarrow 1$ and $\phi\rightarrow 0$ when $r\rightarrow\infty$. 

The simplest potential $V(|\Phi|^2)$ is that of a free field, as given by Eq.~(\ref{e:V_free_field}). It is the only one considered in this section but more complicated examples are given in Sec.~\ref{s:rot}. At the lowest order in $\phi$, all the potentials considered reduce to the free field one so that the following discussions always hold. 
The dominant part of the asymptotic behavior of Eq.~(\ref{e:eqphisph})
with $V(|\Phi|^2)$ replaced by (\ref{e:V_free_field}) is obtained by setting 
$N=1$ and $\Psi=1$:
\be 
\Delta_3 \phi -  \left( \frac{m^2}{\hbar^2}-\omega^2\right) \phi = 0 . 
\ee
For $\omega > m/\hbar$, solutions to this equation are oscillating spherical Bessel functions, which do not decay fast enough to lead to configurations with finite total energy. On the other hand regular solutions for $\omega < m/\hbar$ decay like $\exp\l(-\sqrt{(m/\hbar)^2-\omega^2} \, r\r) /r$ and are admissible solutions to the physical problem. In the following, we will 
focus on this case, i.e. assume $\omega < m/\hbar$. 
When $\omega \rightarrow m/\hbar$, one can anticipate that the field vanishes ($\phi \rightarrow 0$).

\subsection{Numerical code} \label{s:spher_code}

The system (\ref{e:eqpsisph})-(\ref{e:eqphisph}) is solved
by means of a Newton-Raphson iteration implemented in a C++ code built 
on the Kadath library \cite{Grand10,Kadath}. This library enables the use of spectral methods for solving a great variety of partial differential equations that arise in theoretical physics. The 
3-dimensional space
$\Sigma_t$ is decomposed into several numerical domains.
In this particular case spherical shells are used. In the outer numerical domain, space is compactified by making use of the variable $1/r$ so that the computational domain extends up to spatial infinity.
Spectral methods are used \cite{GrandN09}: 
in each domain, the fields are described by their expansions onto a set of known basis functions (typically Chebyshev polynomials). The unknown are then the coefficients of the expansions 
and the resulting non-linear system is solved iteratively.
Regularity near $r=0$ is ensured by using only even Chebyshev polynomials in the numerical domain that encompasses the origin (a more detailed discussion about regularity can be found in the case $k>0$ ; see Sec. \ref{s:solverknotzero}).

\begin{figure}
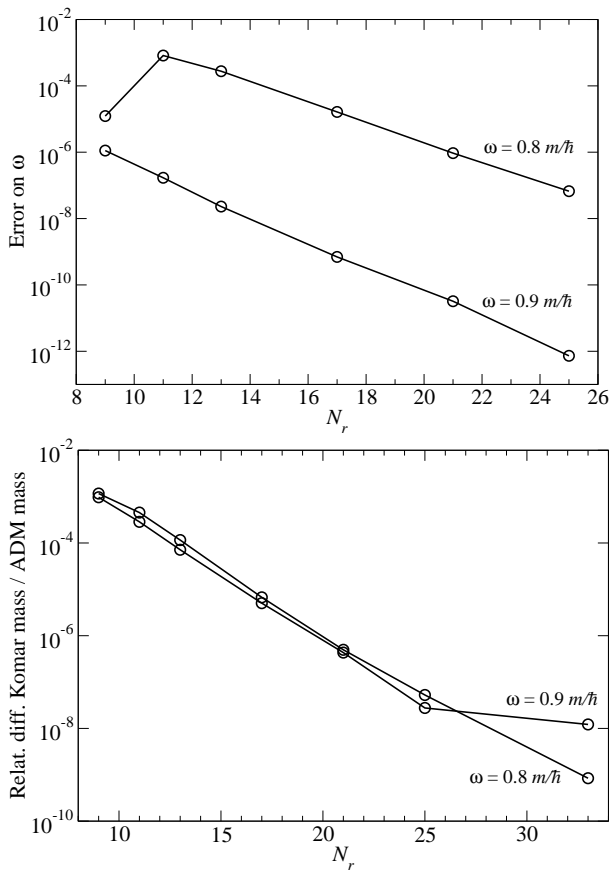

\includegraphics[width=8cm]{conv_ome.eps}
\includegraphics[width=8cm]{error_masses.eps}
\caption{ \label{fig:errsph} For two different values of $\phi_{\rm c}$ (corresponding to $\omega\simeq 0.8 \, m/\hbar$ and $\omega \simeq 0.9 \, m/\hbar$), the top panel shows 
the convergence towards the exact value of $\omega$ (defined as the value found
for $N_r=33$) as a function of $N_r$, while 
the bottom panel shows the difference between the ADM and the Komar masses.}
\end{figure}

Let us point out that an empty flat spacetime ($\phi=0$, $N=1$ and $\Psi=1$) is a trivial solution to the system 
(\ref{e:eqpsisph})-(\ref{e:eqphisph}). In order to avoid convergence to this solution, one demands that the value of the field at the center takes a given non-zero value:
\be
\label{e:centersph}
\phi\l(r=0\r) = \phi_{\rm c}.
\ee
In order to maintain the same number of equations than unknowns, $\omega$ is not treated as a fixed parameter of the solution but rather as an additional unknown. The code finds the appropriate value of $\omega$ so that the condition (\ref{e:centersph}) is fulfilled. This obviously prevents the code from converging to the trivial solution. 

The precision of the code can be assessed by checking the convergence of the value found for $\omega$ when the number of radial coefficients is increased. This is shown on the first panel of Fig.~\ref{fig:errsph}. The convergence is exponential as expected for a well-posed problem solved by spectral methods. 
Another indicator of the code accuracy is the identity between the Komar mass and the ADM mass
of the solution. The Komar mass, expressed above by the volume integral (\ref{e:grav_mass}), 
can be computed from the gradient of the lapse function $N$ on a 2-sphere at spatial 
infinity, while the ADM mass is given by the gradient of the conformal factor $\Psi$
(see e.g. Eq.~(4.15) of Ref.~\cite{Gourg10} for $M_{\rm Komar}$ and
Eq.~(8.48) of Ref.~\cite{Gourg12} for $M_{\rm ADM}$):
\bea
M_{\rm Komar} &=& \frac{1}{4\pi} \lim_{r\rightarrow\infty} \oint_{\mathcal{S}} \partial_r N 
    \, r^2  \sin\theta \, \D\theta \D\varphi \\
\label{e:adm}
M_{\rm ADM} &=& -\frac{1}{2\pi} \lim_{r\rightarrow\infty} \oint_{\mathcal{S}}\partial_r \Psi 
    \, r^2  \sin\theta \, \D\theta \D\varphi , 
\eea
where $\mathcal{S}$ is the 2-sphere of constant coordinate $r$.  
Stationarity implies that $M_{\rm ADM} = M_{\rm Komar} = M$ \cite{Beig78,AshteM79}. This equality can be viewed as a manifestation of the virial theorem \cite{GourgB94}. The second panel of Fig.~\ref{fig:errsph} shows, for two different configurations, the difference between the two masses, as a function of the number of coefficients $N_r$. Once again the difference goes to zero exponentially. The very last point for 
$\omega \simeq 0.9 m/\hbar$ slightly deviates from the exponential behavior probably due to the fact that the Newton-Raphson iteration is stopped at a threshold of $10^{-8}$ or to round-off errors.

\begin{figure}
\includegraphics[width=8cm]{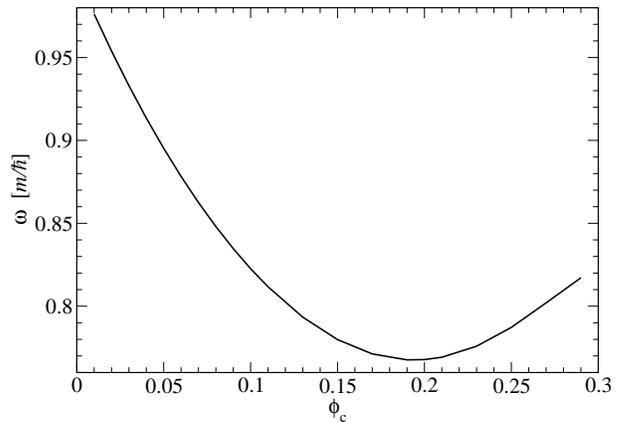}
\caption{ \label{fig:ressph-ome} Value of $\omega$ as a function of $\phi_{\rm c}$ for 
spherical configurations and a free scalar field.}
\end{figure}

\begin{figure}
\includegraphics[width=8cm]{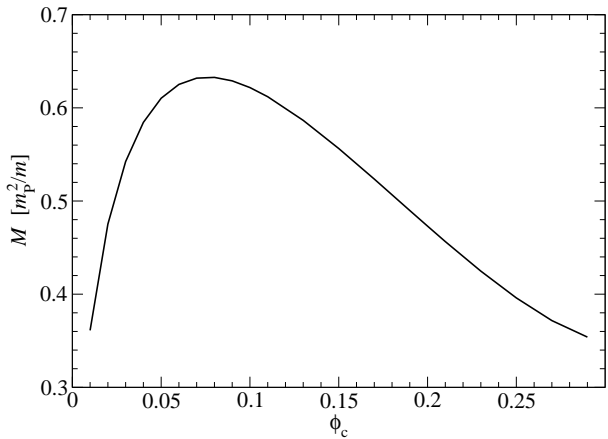}
\caption{ \label{fig:ressph-adm} Gravitational mass $M$ as a function of $\phi_{\rm c}$ for 
spherical configurations and a free scalar field.}
\end{figure}

\begin{figure}
\includegraphics[width=8cm]{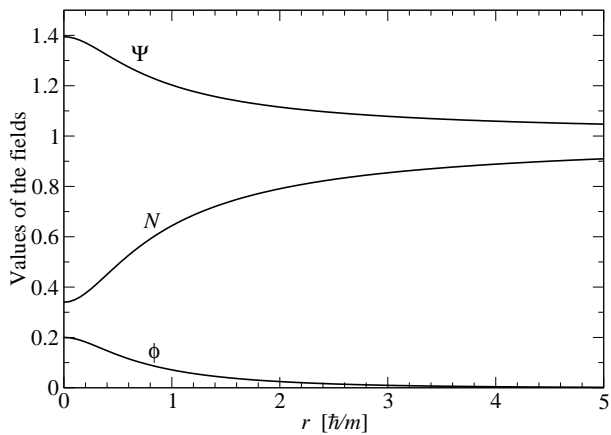}
\caption{ \label{fig:ressph-prof} Radial profiles of the scalar field modulus $\phi$
and of the metric functions $N$ and $\Psi$ for the free-field spherical configuration 
corresponding to $\phi_{\rm c} = 0.2$}
\end{figure}

\subsection{Solutions} \label{s:results_spher}

Some results regarding the spherical case are shown in 
Figs.~\ref{fig:ressph-ome}-\ref{fig:ressph-prof}. The configurations are computed with $N_r=21$ and a decomposition of space in 6 radial domains. Figure~\ref{fig:ressph-ome} shows the value of $\omega$ as a function of $\phi_{\rm c}$. One can see that there is a range of values of $\omega$ for which more than one configuration exist. There is a minimum value of 
$\omega \simeq 0.77 m/\hbar$ for $\phi_{\rm c} \simeq 0.2 $. 
Figure~\ref{fig:ressph-adm} shows the value of the gravitational mass $M$ as a 
function of $\phi_{\rm c}$. We recover the maximum mass found by Kaup \cite{Kaup68}: 
$M_{\rm max} = 0.633 \, m_{\rm P}^2/m$ (reached for $\phi_{\rm c} \simeq 0.07$). Finally 
Fig.~\ref{fig:ressph-prof} shows the radial profiles of $N$, $\Psi$ and $\Phi$ for a configuration close to the minimum value of $\omega$, that is for $\phi_{\rm c} = 0.2$. The results from Fig.~\ref{fig:ressph-ome}-\ref{fig:ressph-adm} confirm the fact that when $\omega \rightarrow m/\hbar$, the fields goes to zero, as does its 
gravitational mass.

\section{Rotating boson stars}\label{s:rot}

\subsection{Equations} \label{s:rot_eq}

In order to construct boson stars that depart from spherical symmetry and 
are rotating, one considers the 
ansatz (\ref{e:ansatzaxe}) with $k\geq 1$.
The obtained configurations are stationary, axisymmetric and \emph{circular},
i.e. the 2-surfaces of transitivity of the spacetime symmetry group 
(the surfaces of constant $(r,\theta)$ in adapted coordinates) are orthogonal
to the surfaces of constant $(t,\varphi)$ \cite{Carte69,Gourg10}.
Thanks to the circularity property, 
we may use \emph{quasi-isotropic coordinates} $(t,r,\theta,\varphi)$ (also called 
\emph{Lewis-Papapetrou coordinates}), in 
which $\beta^i=(0,0,\beta^\varphi)$ and $\gamma_{ij} = \mathrm{diag}(A^2, A^2 r^2, B^2 r^2\sin^2\theta)$, 
so that the 4-dimensional metric line element reads (see e.g. \cite{Gourg10,FriedS13})
\bea
 g_{\mu\nu} \, \D x^\mu \, \D x^\nu   & = & - N^2 \D t^2 + A^2 \left(
    \D r^2 + r^2 \D\theta^2 \right) \nonumber \\
        & & + B^2 r^2 \sin^2\theta \left(\D\varphi + \beta^\varphi \D t\right)^2. 
                \label{e:qimet}
\eea 
The metric is thus entirely described by four functions of $(r,\theta)$:
$N$, $A$, $B$ and $\beta^\varphi$. Note that the spherically symmetric case 
treated in Sec.~\ref{s:spheric} is recovered for $A=B=\Psi^2$ and $\beta^\varphi=0$. 

\begin{widetext}

The Einstein equation (\ref{e:Einstein_eq}) leads to the following system
(see e.g. \cite{Gourg10,FriedS13}, taking into account that these references 
make use of $\omega=- \beta^\varphi$):
\bea
\Delta_3 \nu &= &4\pi A^2 \l(E+S\r) + \frac{B^2r^2\sin^2\theta}{2N^2} \partial \beta^\varphi \partial \beta^\varphi  - \partial\nu \partial\l(\nu + \ln B\r) \label{e:nu} \\
\tilde{\Delta}_3\l(\beta^\varphi r \sin\theta\r)& = & 16 \pi \frac{N A^2}{B^2} \frac{P_\varphi}{r\sin\theta}   + r\sin\theta\partial \beta^\varphi\partial\l(\nu - 3\ln B\r) \label{e:Nphi} \\
\Delta_2 \l[\l(NB -1\r)r \sin\theta\r] & = &  8 \pi N A^2 B r \sin\theta \l(S^r_{\ \, r} + 
    S^\theta_{\ \, \theta} \r) 
    \label{e:B} \\
\Delta_2\l(\ln A  + \nu\r) & = & 8 \pi A^2 S^\varphi_{\ \, \varphi} + \frac{3B^2 r^2\sin^2\theta}{4N^2} \partial \beta^\varphi \partial \beta^\varphi - \partial \nu \partial \nu ,  \label{e:A}
\eea
where $\nu := \ln N$ and
\bea
\Delta_3 &:=& \frac{\partial^2}{\partial r^2} + \frac{2}{r}\frac{\partial}{\partial r} + \frac{1}{r^2} \frac{\partial^2}{\partial\theta^2} + \frac{1}{r^2 \tan \theta}\frac{\partial}{\partial \theta} \\
\tilde{\Delta}_3 &:=& \Delta_3 - \frac{1}{r^2\sin^2\theta} \\
\Delta_2 &:=& \frac{\partial^2}{\partial r^2} + \frac{1}{r}\frac{\partial}{\partial r} + \frac{1}{r^2} \frac{\partial^2}{\partial\theta^2} \\
\partial f \partial g &:=& \frac{\partial f}{\partial r} \frac{\partial g}{\partial r} + \frac{1}{r^2}\frac{\partial f}{\partial\theta}\frac{\partial g}{\partial\theta}.
\eea
$\Delta_3$ (resp. $\Delta_2$) is  the 3-dimensional flat Laplacian (resp. 2-dimensional) applied to axisymmetric functions. $\partial f \partial g$ denotes the scalar product of the gradients of $f$ and $g$, with respect to the flat metric.
The quantities $E$, $S$, $S^i_{\ \, j}$ and $P_\varphi$ are related to the 3+1 decomposition of the energy-momentum tensor, as defined in Appendix~\ref{s:app_T}. 
Using  Eqs.~(\ref{e:E_axisym})-(\ref{e:S_axisym}), we
find the explicit form of the terms involved in the right-hand side of (\ref{e:nu})-(\ref{e:A}):
\bea
E + S &=& \frac{2}{N^2} (\omega + k\beta^\varphi)^2 \phi^2 - V \label{e:EpS} \\
P_\varphi &=& \frac{k}{N} (\omega + k \beta^\varphi) \phi^2 \label{e:Pphi} \\
S^r_{\ \, r} + S^\theta_{\ \, \theta} & = &  \left[ 
    \frac{1}{N^2} (\omega + k\beta^\varphi)^2 - \frac{k^2}{B^2 r^2 \sin^2\theta} \right]
    \phi^2  - V \label{e:Srr_Sthth}\\
S^\varphi_{\ \, \varphi} & = & \frac{1}{2} \left\{ 
    \left[ \frac{1}{N^2} (\omega + k\beta^\varphi)^2 + \frac{k^2}{B^2 r^2 \sin^2\theta} \right]
    \phi^2 - \frac{1}{A^2} \partial \phi \partial \phi - V \right\} 
\eea

The scalar field $\Phi$ obeys the Klein-Gordon equation (\ref{e:kg}), which
becomes, once the metric (\ref{e:qimet}) and the ansatz (\ref{e:ansatzaxe}) are used,
\be
\label{e:kgaxe}
\Delta_3 \phi - \frac{k^2 \phi}{r^2\sin^2\theta} = A^2 \left[
    \frac{{\rm d}V}{{\rm d}\l|\Phi\r|^2} - \frac{1}{N^2} (\omega + k \beta^\varphi)^2 \right]
     \phi - \partial \phi \partial \l(\nu + \ln B\r) + \l(\frac{A^2}{B^2}-1\r) \frac{k^2\phi}{r^2\sin^2\theta}. 
\ee
The system of equations is closed by demanding that one recovers empty flat spacetime at infinity, i.e. that $N\rightarrow 1$, $A\rightarrow 1$, $B\rightarrow 1$, $\beta^\varphi\rightarrow 0$ and $\phi\rightarrow 0$ when $r\rightarrow \infty$.
\end{widetext}

The main difference between the spherical and axisymmetric boson stars is a change in the topology of the field $\phi$. Indeed, for regularity reasons, $\phi$ must vanish on the rotation axis
($\theta=0$ or $\theta=\pi$). It follows that the shape of the scalar field is no longer spherical but rather toroidal. More precisely, close to the axis, the field behaves like $\l(r \sin\theta\r)^k$. For $k\geq 2 $ this ensures regularity on the axis because all the divisions by $\sin^2\theta$ that appear 
in Eqs.~(\ref{e:Srr_Sthth})-(\ref{e:kgaxe}) are made possible. The case $k=1$ is slightly more subtle, especially concerning Eq.~(\ref{e:kgaxe}), for terms like $\displaystyle\frac{\phi} {\sin^2\theta}$ may appear singular at first glance. However the potentially singular part in the left-hand side of Eq.~(\ref{e:kgaxe}) is $ \displaystyle\frac{1}{\tan\theta}\displaystyle\frac{\partial \phi}{\partial \theta} -\displaystyle\frac{\phi} {\sin^2\theta}$ and a direct computation enables to verify that the singularities cancel. On the right-hand side of Eq.~(\ref{e:kgaxe}), the term  $\l(\displaystyle\frac{A^2}{B^2}-1\r) \displaystyle\frac{k^2\phi}{\sin^2\theta}$ may seem problematic when $\phi$ vanishes only like $\sin\theta$. However it is known that on the axis $A = B$ (local flatness, cf. \cite{Gourg10})
so that the term $\l(\displaystyle\frac{A^2}{B^2}-1\r)$ vanishes at least as $\sin\theta$, thus ensuring regularity, even in the $k=1$ case. The regularity near the origin $r=0$ is ensured by the basis decompositions used in the innermost domain, as discussed in \ref{s:solverknotzero}.

\subsection{Spectral solver}\label{s:solverknotzero}

Let us note that boson stars have some common features with a class of field solutions known as 
\emph{vortons} and already studied by means of Kadath in a previous article \cite{GrandFF11}. 
In the vorton case, a complex field $\sigma$ has the same geometry as the $\Phi$ field. However, instead of being coupled to gravity there is a second complex field and the two interacts. Nevertheless, the numerical treatment of the vorton field $\sigma$ and the boson field $\Phi$ is very similar.

The axisymmetric boson stars are computed using the polar space of the Kadath library, where fields are given in terms of the $\l(r, \theta\r)$ coordinates. As usual, space is divided into several radial domains and extends up to spatial infinity. Real scalar fields, like the lapse $N$, are expanded onto even cosines with respect to the angular variable $\theta$. As far as the radial coordinate is concerned, standard Chebyshev are employed, except in the domain that encompasses the origin, for which only even Chebyshev polynomials are used. We will refer to such basis of decomposition as the {\em even basis}. Another basis is the {\em odd basis}, where odd sines are used with respect to $\theta$ and odd Chebyshev polynomials with respect to $r$, near the origin. It is for instance easy to see that if a field $f$ is expanded onto the even basis, then a ratio like $f / \l(r\sin\theta\r)$ must be expanded onto the odd one. Let us mention that, due to the non-local nature of spectral methods, the use of those basis is valid throughout the innermost domain, no matter what its size is.

The metric fields $N$, $A$, $B$ and $\beta^\varphi$ must be expanded onto the even basis. This can be understood by noting that $\D s^2 = g_{\mu\nu}\D x^\mu \D x^\nu$ in Eq.~(\ref{e:qimet}) must be a scalar field. The case of $\phi$ is different in the sense that the ``true'' scalar is the field $\Phi$ itself. In other words, $\phi$ is only the harmonic $k$ of a genuine scalar field. It follows that $\phi$ must be expanded onto the even basis if $k$ is even but onto the odd one when $k$ is odd. This situation is the same as that for the $\sigma$ field in the vorton case \cite{GrandFF11}. This choice of decomposition is also consistent with the regularity condition that $\phi$ must vanish like $\l(r\sin\theta\r)^k$ on the axis. For completeness, let us mention that the numerical code does not search directly for the fields $N$, $B$, $A$ and $\beta^\varphi$ but rather works with the auxiliary fields appearing on the left-hand side of Eqs.~(\ref{e:nu})-(\ref{e:A}) which are $\nu$, $\beta^\varphi r \sin\theta$, $\l(NB -1\r)r \sin\theta$ and $\ln A  + \nu$.

The system is solved by means of Newton-Raphson iterative scheme. 
For each value of $k$, the most difficult part consists in finding a first solution. Once this is done, $\omega$ can be slowly changed to construct the whole family of configurations. This is to be contrasted with the parameter $k$, which is an integer and so cannot be modified in this manner. For each $k$ we proceed as follows. First we consider an initial guess for $\phi$ of the form 
\be
\label{e:initial}
\phi \l(r, \theta\r) = f_0  \l(r\sin\theta\r)^k \exp\l(-x^2/\sigma_x\r) \exp \l(-z^2/\sigma_z\r) , 
\ee
where $x:=r\sin\theta$, $z:= r\cos\theta$, 
and $f_0$, $\sigma_x$ and $\sigma_y$ are three constants that can be freely chosen; 
they control the amplitude of the field and the shape of the toroidal configuration. 
The form (\ref{e:initial}) ensures that the regularity condition on the axis is fulfilled and that the field decays as expected at spatial infinity.
As for the spherical case (cf. Sec.~\ref{s:spher_code}), in
order to avoid that the solver converges to the trivial solution $\phi = 0$,  $\omega$ is treated as an unknown and one demands that the field takes a given non-zero value $\phi_0$ at some point $(r,\theta)=(r_0,\pi/2)$
in the equatorial plane (we have to choose $r_0 \not= 0$ since $\phi(r=0)=0$ for $k\geq 1$). In order to facilitate convergence, one also tries to work in cases where the scalar field amplitude is small and the metric close to Minkowski spacetime. As already stated, this should give a value of $\omega$ close to $m/\hbar$. After a few trials, it is usually possible to find a choice of $f_0$, $\sigma_x$, $\sigma_y$, $r_0$ and $\phi_0$ that leads to an admissible solution, i.e. that converges to a solution with $\omega < m/\hbar$. For instance, choosing 
$r_0=35 \, \hbar/m$, $\phi_0 = 0.001$, $f_0=5\cdot 10^{-9} \hbar^4/m^4$, 
$\sigma_x= 612\, \hbar^2/m^2$ and $\sigma_z = 306\, \hbar^2/m^2$ proved to be a valid choice for $k=4$. Once again, this is done only once for each value of $k$, the other solutions being found by slowly varying $\omega$.

\subsection{Error indicators}\label{ss:errors}

In order to check the accuracy of the code, several error indicators can be defined. First, as in the spherical case
(Sec.~\ref{s:spher_code}), one can check whether the ADM 
and Komar expressions of the gravitational mass $M$ agree.

The second error indicator was first obtained by Bonazzola \cite{Bonaz73} and arises because of the presence of a 2-dimensional Laplace operator in Eq.~(\ref{e:A}). Using the associated Green function, one can show that the solution decreases fast enough if, and only if, the source term (i.e. the right-hand side of Eq.~(\ref{e:A})) has no 2-dimensional monopolar contribution. This is equivalent to
\bea
&& I := \int_{r=0}^\infty \int_{\theta=0}^\pi \Bigg[ \pi A^2 S^\varphi_{\ \, \varphi} + \frac{3B^2 r^2\sin^2\theta}{4N^2} \partial \beta^\varphi \partial \beta^\varphi \nonumber \\
&& \qquad\qquad\qquad\qquad\qquad\qquad
 - \partial \nu \partial \nu\Bigg] r {\rm d}r {\rm d}\theta = 0. \label{e:grv2}
\eea
This is the so-called GRV2 identity \cite{BonazG94}. Let us mention that even if Eq.~(\ref{e:B}) does involve another $\Delta_2$ operator, it does not lead to such condition. Indeed, the source being proportional to $r\sin\theta$, it has, by construction, no monopolar term. In previous works,
for instance in the \texttt{Lorene/nrotstar} code (see Appendix~B of \cite{Gourg10}), it 
was necessary to enforce the condition $I=0$ at each step of the iteration by modifying the source of 
Eq.~(\ref{e:A}). With Kadath, no such treatment is required. This is probably due to the fact that the system is solved as a whole and not by separating the various equations in terms of operators on one side and source on the other one.

The third error indicator regards the computation of the angular momentum $J$
introduced in Sec.~\ref{s:global}. It can be evaluated by means of the 
volume integral (\ref{e:J_vol_int}) with $\sqrt{\gamma} \, \D^3 x = A^2 B r^2\sin\theta \, \D r\, 
\D \theta\, \D \varphi$ in quasi-isotropic coordinates. 
Let us call $J_{\rm v}$ this value. 
An alternative way to compute $J$ is via the Komar surface integral, which can be written as 
(see e.g. Sec.~4.4 of Ref.~\cite{Gourg10})
\be
\label{e:jinf}
J =  \frac{1}{16\pi} \lim_{r\rightarrow \infty} \oint_{\mathcal{S}} \partial_r \beta^\varphi r^4 \sin^3\theta 
    \, \D\theta\, \D\varphi , 
\ee
where $\mathcal{S}$ is the 2-sphere of coordinate radius $r$.
Let us call $J_{\rm s}$ the numerical value of $J$ hence obtained. 
Note that thanks to the compactification of the last domain, $r=+\infty$ belongs
to the computational domain and both expressions can be computed directly. 
One can then check whether $J_{\rm v} = J_{\rm s}$.

\begin{figure}[!hbtp]
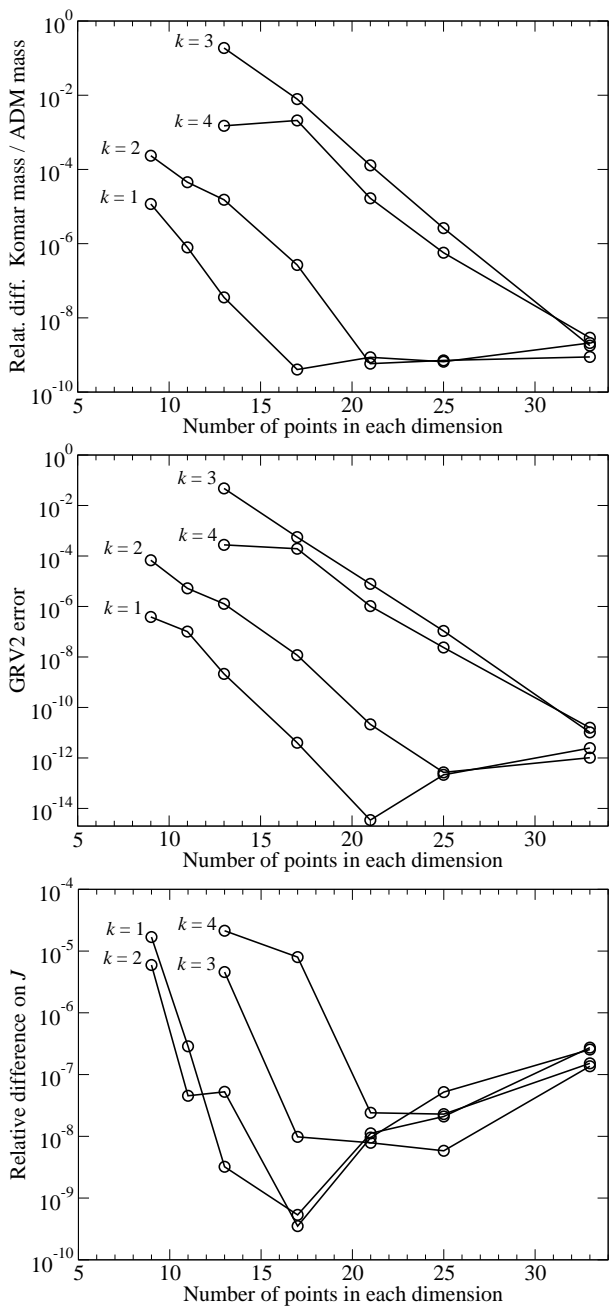

\includegraphics[width=8cm]{err_masses.eps}
\includegraphics[width=8cm]{err_grv2.eps}
\includegraphics[width=8cm]{err_j.eps}
\caption{ \label{fig:errors}
Various error indicators as the number of spectral coefficients in both the radial and angular dimensions, for free scalar field configurations with $\omega=0.8\, m/\hbar$ 
and rotational quantum number $k=1, 2, 3$ and $4$. The first panel shows the error on the masses, the second one on the GRV2 identity and the last one on the angular momentum.
}
\end{figure}

The error indicators are shown on the different panels of Fig.~\ref{fig:errors} for different spectral resolutions. By resolution, one means the number of points in both the radial and the angular dimensions (those numbers are thereafter kept identical). Results are shown for $\omega=0.8\, m/\hbar$ and $k=1, 2, 3$ and $4$. More precisely, the first panel shows the error on the masses defined as $\l|M_{\rm Komar} - M_{\rm ADM}\r| / \l|M_{\rm Komar} + M_{\rm ADM}\r|$. On the second panel the error on the GRV2 identity is plotted and defined as being $\l|I_{\Phi} + I_{\rm grav}\r| / \l|I_{\Phi} - I_{\rm grav}\r|$, where $I_{\Phi}$ is the part of Eq.~(\ref{e:grv2}) that contains the $S^\varphi_{\ \, \varphi}$ term and $ I_{\rm grav}$ the remaining terms. The third panel of Fig.~\ref{fig:errors} shows the relative difference 
$\l|J_{\rm s} - J_{\rm v}\r| / \l|J_{\rm s} + J_{\rm v}\r|$
of the two expressions for the angular momentum. All the error indicators exhibit a similar behavior, that is a spectral convergence \cite{GrandN09}: the error decreases exponentially  and then saturates due to round-off errors. The error on the angular momentum seems to slightly increase at very high resolution. By looking separately at the convergence of $J_{\rm s}$ and $J_{\rm v}$ with the resolution, one can see that the error is dominated by the surface integral $J_{\rm s}$. In this case, the round-off errors are greater because of the multiplication by $r^4$ that appears in Eq.~(\ref{e:jinf}). Those errors accumulate which explains the increase of the error at very high resolution. This suggests that it is preferable to use 
the volume integral (\ref{e:J_vol_int}) to compute the angular momentum.

\subsection{Numerical results for a free scalar field} \label{s:resu_free_field}

In this section numerical results for rotating 
boson stars with $k=1$, $2$, $3$ and $4$ and the free-field potential (\ref{e:V_free_field}) are presented. The same numerical setting is used for almost all the computations. It consists in a decomposition of the 3-dimensional space $\Sigma_t$ into 8 spherical domains. The last compactified domain extends from 
$r=64\, \hbar/m$ up to infinity. In each domain, 21 coefficients are used for both coordinates $r$ and $\theta$. For the most relativistic configurations in the cases $k=3$ and $k=4$, up to 33 coefficients are used. The Newton-Raphson iteration is stopped at the threshold of $10^{-8}$. This setting has been chosen to ensure a good accuracy in all the quantities presented thereafter. However, in some cases, this is not the best possible choice. For instance, in the case $k=4$, the boson stars have a much larger size than in the $k=1$ case and would benefit from using more extended domains. An extensive survey of the parameter space would require some fine-tuning of the numerical parameters to ensure convergence and is beyond the scope of this paper. 

\begin{figure}[!hbtp]
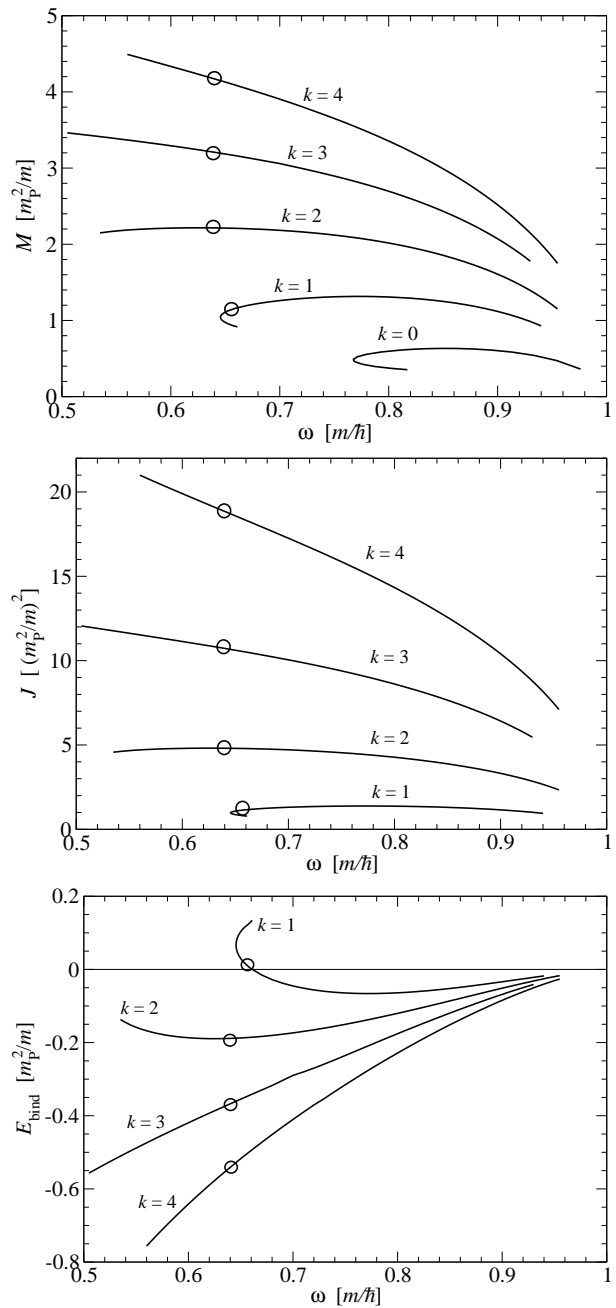

\includegraphics[width=8cm]{adm_axe.eps}\\[1ex]
\includegraphics[width=8cm]{moment.eps}\\[1ex]
\includegraphics[width=8cm]{eb.eps}
\caption{ \label{fig:global}
Global quantities as functions of $\omega$, 
namely the gravitational mass (first panel), the angular momentum (second panel) and the binding energy (third panel). Configurations at the left side of the circles possess an ergoregion and thus may be unstable (see Sec.~\ref{s:ergo}).
}
\end{figure}

Global quantities are plotted in Fig.~\ref{fig:global}. The gravitational mass [Eq. (\ref{e:grav_mass})] and the total angular momentum [Eq.~(\ref{e:J_vol_int})] are plotted as functions of $\omega$. The binding energy is also shown; it is defined by 
\be
\label{e:binding}
E_{\rm bind} = M - \mathcal{N} m,
\ee
with the particle number $\mathcal{N}$ given by (\ref{e:Q_3p1}). 
Maximum mass configurations are observed for $k=0$, $k=1$ and $k=2$. 
The $k=0$ case is the Kaup limit $M_{\rm max}^{(k=0)} = 0.633 \, m_{\rm P}^2/m$ 
discussed in Sec.~\ref{s:results_spher}.
For $k=1$, we recover the value found by Yoshida and Eriguchi \cite{YoshiE97b}:
\be \label{e:M_max_k_1}
    M_{\rm max}^{(k=1)} = 1.315 \, \frac{m_{\rm P}^2}{m} . 
\ee
The maximum mass for $k=2$ could not be determined in Ref.~\cite{YoshiE97b},
but the lower bound $M_{\rm max}^{(k=2)} \geq 2.21\,  m_{\rm P}^2/m$ was
established, with the hint that the maximum mass was not far from it
(cf. Fig.~3 in Ref.~\cite{YoshiE97b}). In agreement with this lower bound, 
we find here (cf Fig.~\ref{fig:global})
\be
    M_{\rm max}^{(k=2)} = 2.216\,  \frac{m_{\rm P}^2}{m} .
\ee
There is little doubt that maximum masses also exist for $k\geq 3$ but 
these configurations are difficult to get given our standard numerical setting. 
Note however that if such maximum mass stars exist, they have an ergoregion 
as shown in Fig.~\ref{fig:global} and therefore are likely to be unstable
(see Sec.~\ref{s:ergo}).  

In the case $k=1$, a turning point is observed around $\omega = 0.64 \, m/\hbar$, meaning that no solutions are found for smaller values of $\omega$. This also implies that there are values of $\omega$ for which two different boson stars coexist. For $k=1$, configurations with positive binding energy are observed;
they are expected to be unstable.

\begin{figure}[!hbtp]
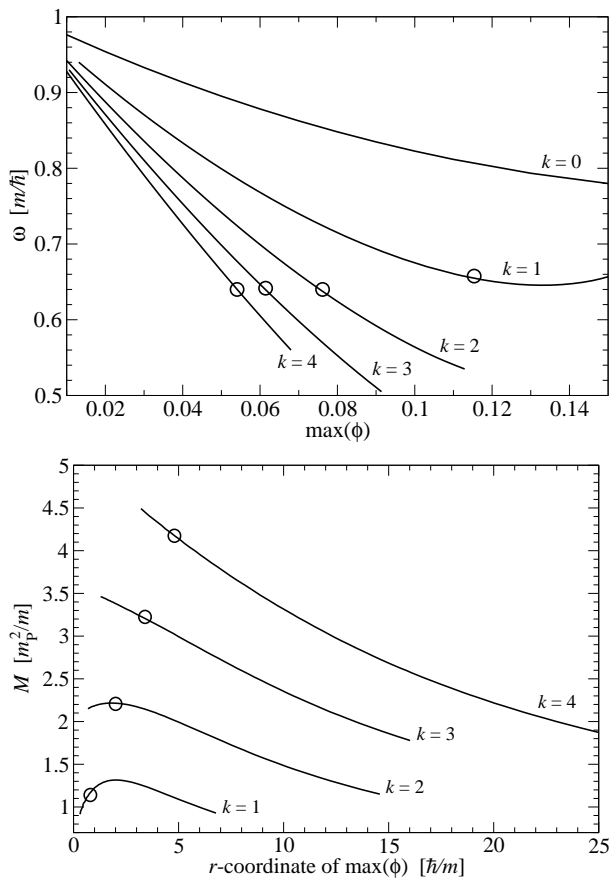

\includegraphics[width=8cm]{max.eps} \\[1ex]
\includegraphics[width=8cm]{posmax.eps}
\caption{ \label{fig:max}
The first panel shows the value of $\omega$ as a function of the maximum value of the 
scalar field modulus $\phi$, whereas the second one shows the gravitational 
mass as a function of the radius at which this maximum is attained. As in Fig.~\ref{fig:global}, circles denote the configurations for which ergoregions start to appear.
}
\end{figure}

\begin{figure}[!hbtp]
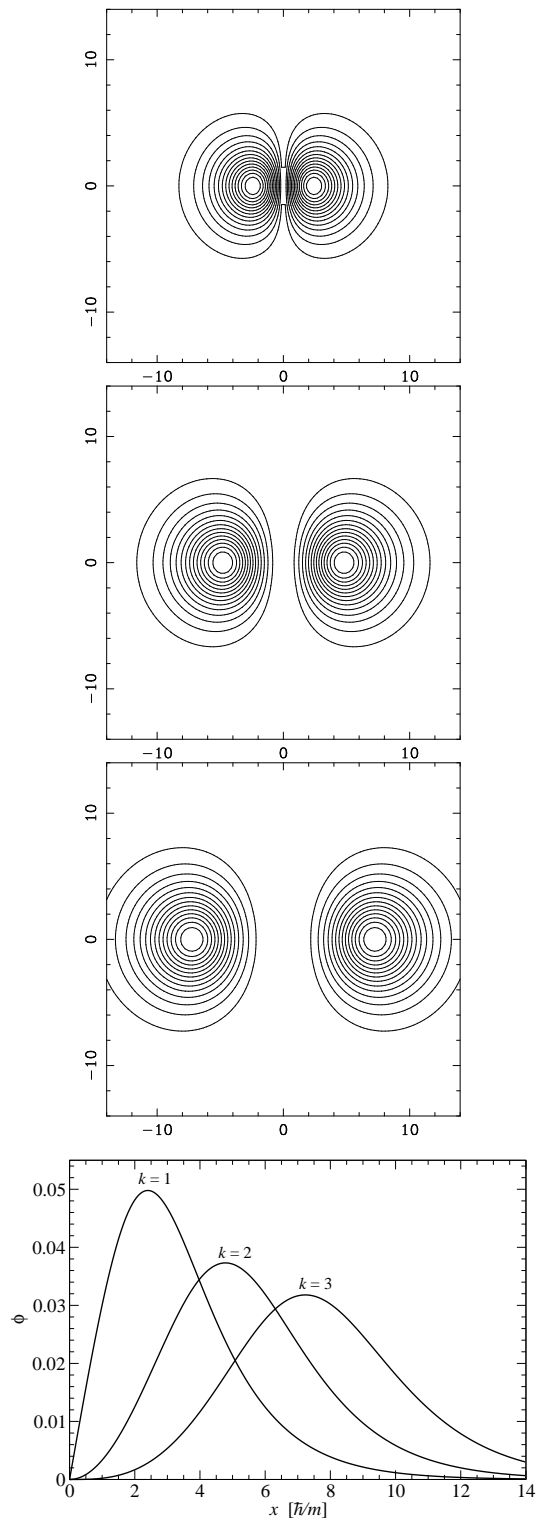

\includegraphics[width=5cm]{phi1.eps}
\includegraphics[width=5cm]{phi2.eps}
\includegraphics[width=5cm]{phi3.eps}\\[2ex]
\includegraphics[width=7cm]{prof_0.8.eps}
\caption{ \label{fig:contours}
Isocontours of the scalar field modulus $\phi$ in a meridian plane of
constant $(t,\varphi)$ 
for $\omega=0.8\, m/\hbar$ in all plots and $k=1$ (first panel), $k=2$ (second one) and $k=3$ (third one).
The Cartesian-like coordinates used for the plot are $x:=r\sin\theta$ (horizontal axis) 
and $z:=r\cos\theta$ (vertical axis), in units of $\hbar/m$.
The fourth panel shows, for the same configurations, the profiles of $\phi$ along the $x$-axis,
i.e. the functions $\phi(r,\pi/2)$. 
}
\end{figure}

Figure~\ref{fig:max} shows quantities related to the value of the scalar field $\Phi$. More precisely, the first panel shows $\omega$ as a function of the maximum value of $\Phi$ and the second panel shows the gravitational mass $M$ as a function of the radius $r_{\rm max}$ at which the maximum value of $\Phi$ is attained. This last plot illustrates clearly the fact that the boson stars size increases with $k$. This effect is also seen on Fig.~\ref{fig:contours} where isocontours of the scalar field are shown, for $k=1$, $2$ and $3$, and for a fixed value of $\omega$ ($0.8\, m/\hbar$). Fig.~\ref{fig:contours}  also shows (last panel) the corresponding profiles of the scalar field along the $x$-axis.

\begin{figure}[!hbtp]
\includegraphics[width=5cm]{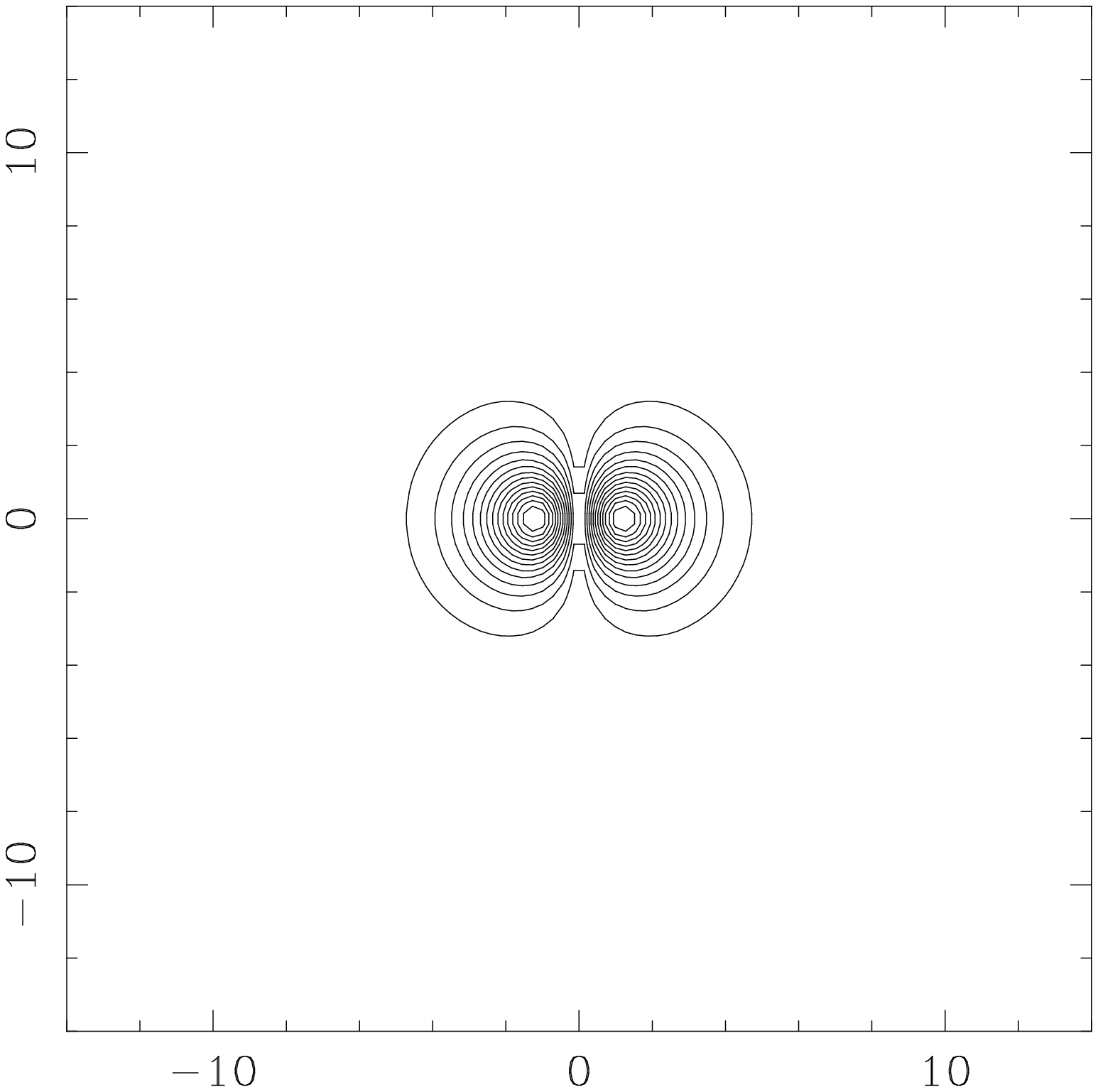}
\includegraphics[width=5cm]{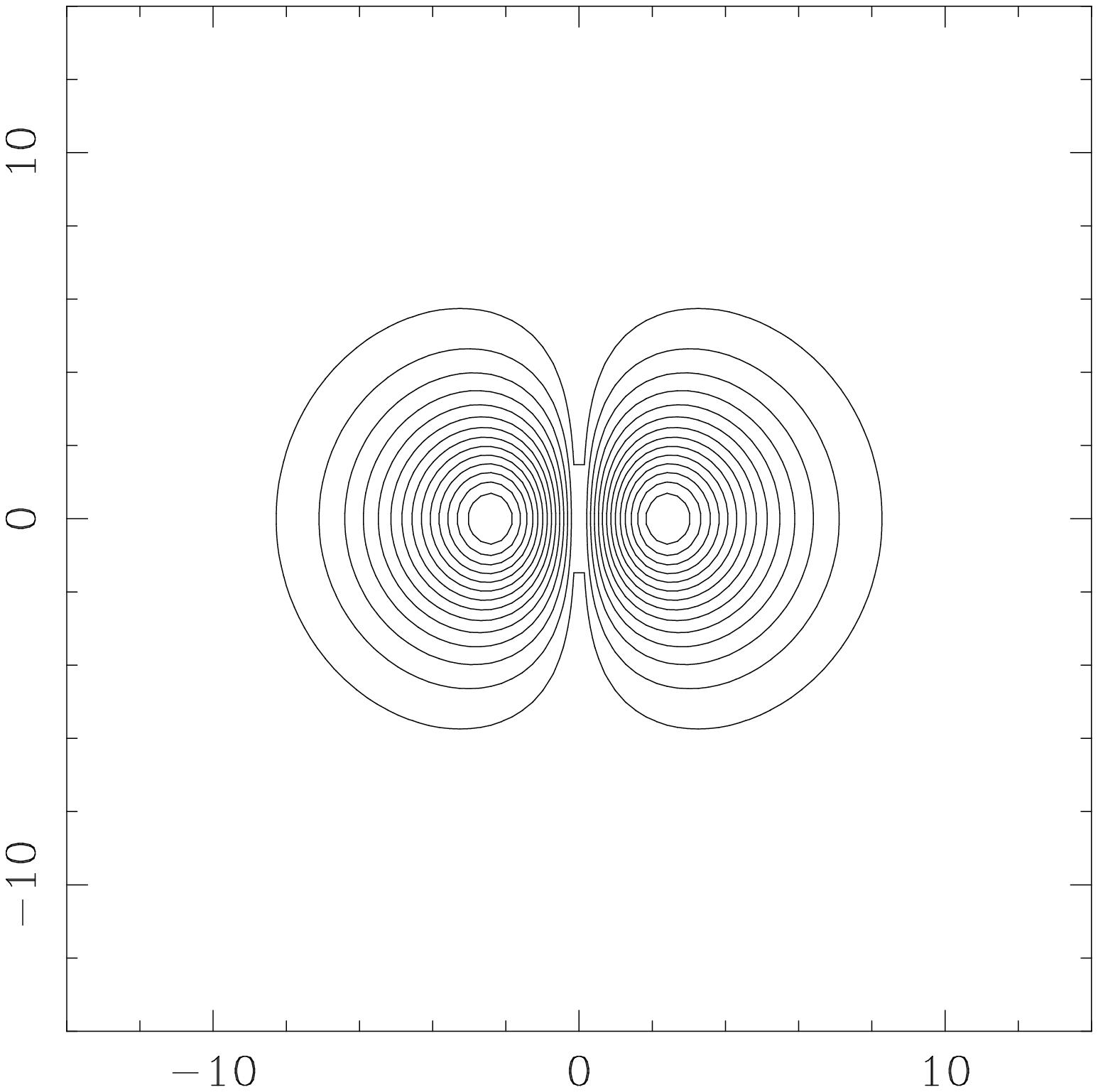}
\includegraphics[width=5cm]{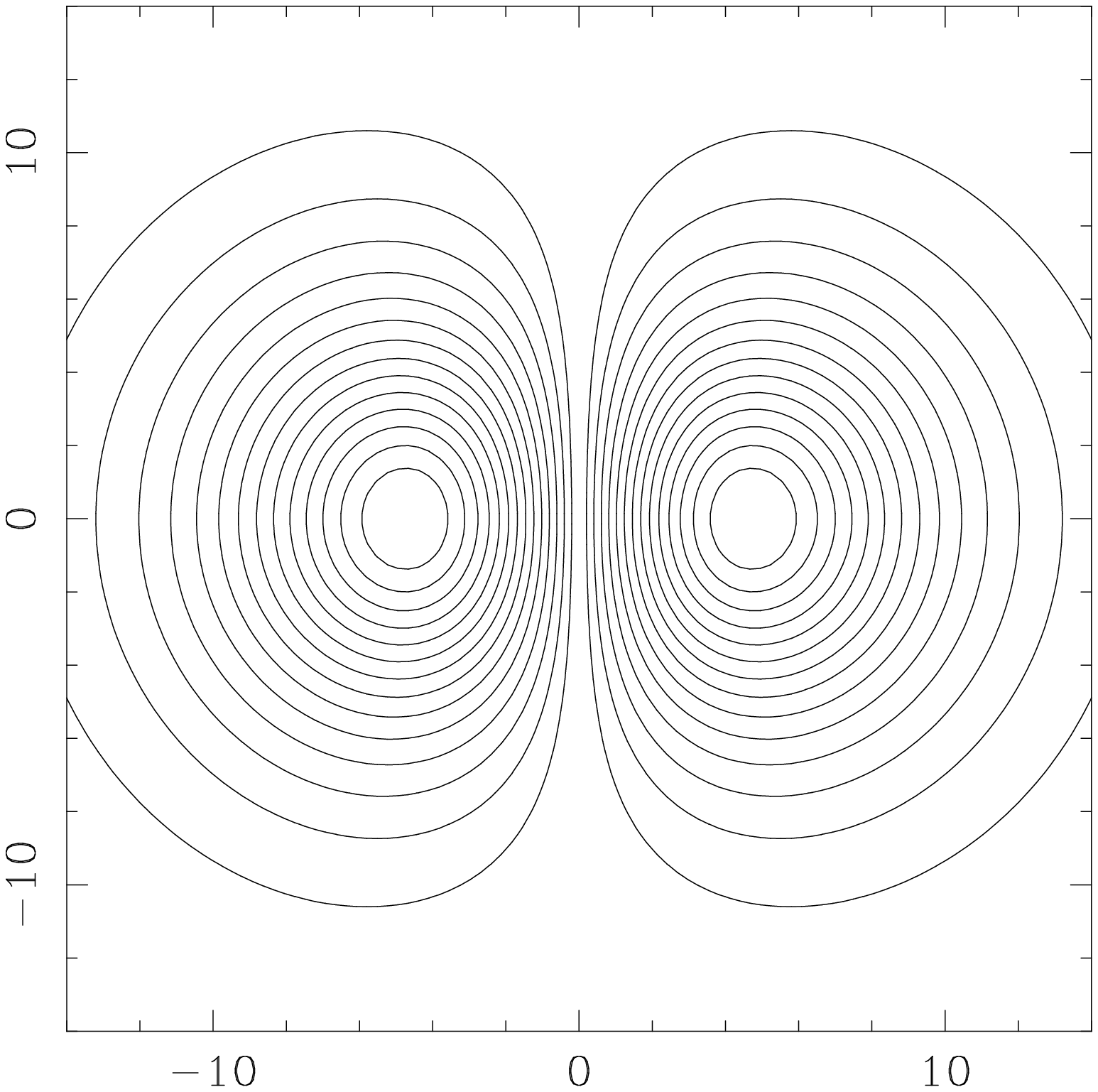}\\[2ex]
\includegraphics[width=7cm]{prof_k1.eps}
\caption{ \label{fig:contours_ome}
Same as Fig.~\ref{fig:contours} but for $k=1$ in all plots and  
$\omega=0.7\, m/\hbar$ (first panel), $\omega=0.8 \, m/\hbar$ (second one) and $\omega=0.9\, m/\hbar$ (third one).
}
\end{figure}

Figure~\ref{fig:contours_ome} shows the effect of $\omega$ on the structure of the scalar field. 
It illustrates the fact that the configurations are more and more extended as $\omega$ 
approaches $m/\hbar$.

\subsection{Ergoregions} \label{s:ergo}

A highly relativistic rapidly rotating object can develop an \emph{ergoregion}, i.e. a spacetime 
region where the Killing vector $\partial/\partial t$ becomes spacelike; in more physical terms, this
means that, in such a region, no observer can remain static with respect to asymptotically inertial observers, due to some strong frame dragging effect. 
This concept is well known for a Kerr black hole, for which, as long as the angular momentum 
differs from zero, an ergoregion exists outside the event horizon. 
The existence of ergoregions in some rotating boson star models has been
demonstrated by Kleihaus et al.~\cite{KleihKLS08}.

As found by Friedman \cite{Fried78}, scalar field configurations
are unstable in spacetimes with an ergoregion but no event horizon, 
the instability mechanism being linked to superradiant scattering. 
The timescale of the instability depends on the spherical harmonic azimuthal
index $m$ of the perturbation: it is very large for $m\gg 1$
\cite{CominS78} and smaller for $m\sim 1$ \cite{YoshiE96}. 
Some authors have put forward the existence of ergoregions to eliminate
boson stars as viable alternatives to black holes in rapidly rotating compact
objects \cite{CardoPCC08}. 
However no exact computation of the instability timescale for
rotating boson stars has been performed yet. Accordingly, we shall consider
configurations with an ergoregion as \emph{potentially} ruled out from an 
astrophysical viewpoint, their exact status depending whether the 
instability timescale is larger or shorter than the age of the Universe. 

\begin{figure}[!hbtp]
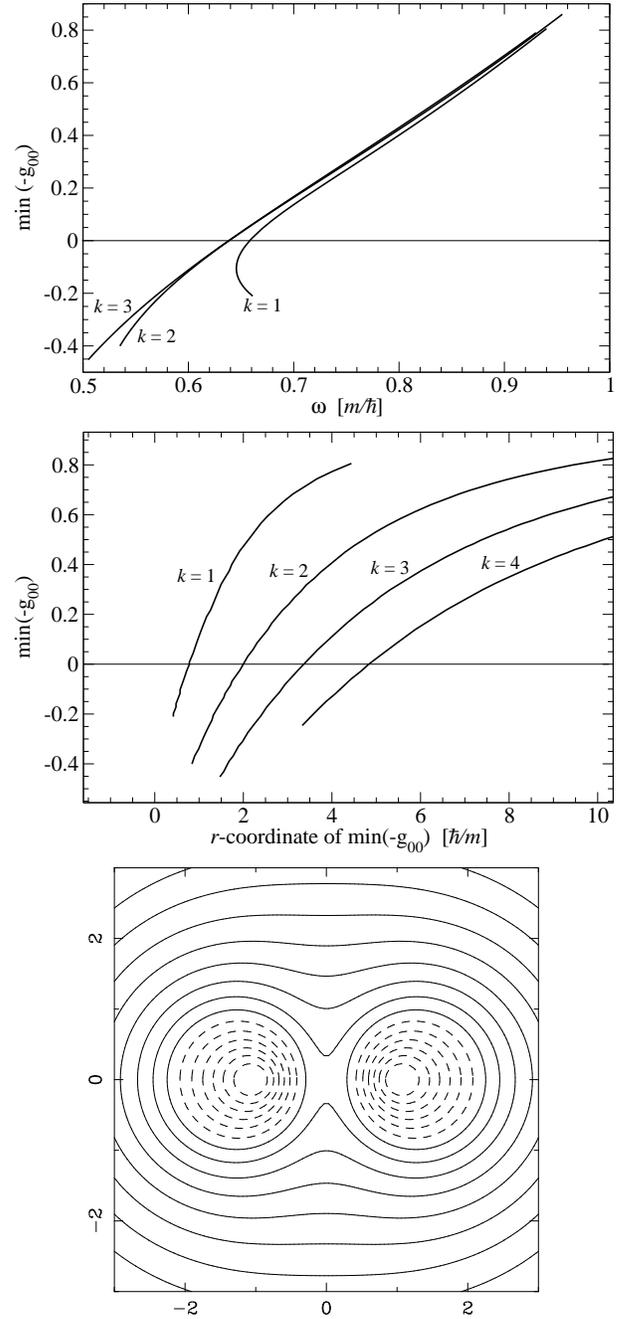

\includegraphics[width=8cm]{ergo_ome.eps}\\[1ex]
\includegraphics[width=8cm]{ergoindic.eps}\\[1ex]
\includegraphics[width=6cm]{ergo.eps}
\caption{ \label{fig:ergo}
The first and second panel shows the value of the minimum of $-g_{00}$ for free-field boson stars 
of different rotational quantum number $k$, 
as a function of $\omega$ and the location of the minimum, respectively. Isocontours of $-g_{00}$ 
in a plane of constant $(t,\varphi)$ are plotted in the third panel, for $k=2$ and $\omega=0.55\, m/\hbar$. The Cartesian-like coordinates of this plot are $x:=r\sin\theta$ and $z:=r\cos\theta$,
in units of $\hbar/m$. The ergoregion is the torus shown in dashed-lines.
}
\end{figure}

The assumption that $\partial/\partial t$ is spacelike (ergoregion definition)
is equivalent to $g_{00} > 0$ or, in view 
of the quasi-isotropic line element (\ref{e:qimet}), to
\be \label{e:ergo_criterion}
    - g_{00} = N^2 - \l(B \beta^\varphi\r)^2 r^2 \sin^2 \theta < 0 .
\ee 
The region defined by the above equation is topologically a toroid. 
The minimal value of $-g_{00}$ is plotted in Fig.~\ref{fig:ergo} for all the 
rotating configurations. When it is negative, an ergoregion exists. 
The first panel shows that ergoregions appear at approximately the same value of $\omega$: 
$\omega \simeq 0.66\, m/\hbar$ for $k=1$ and $\omega \simeq 0.64\, m/\hbar$ for $k=2$, 
$k=3$ and $k=4$, the last two being indistinguishable.
For $\omega$ larger than this critical value, the configurations are not sufficiently relativistic
for an ergoregion to exist. Once again, the second panel shows that boson stars are more extended for higher values of $k$, as can be seen with the increase of the radius of the minimum with $k$. The third panel of Fig.~\ref{fig:ergo} shows the isocontours of $-g_{00}$
in a meridional plane of constant $(t,\varphi)$, for $k=2$ and $\omega=0.55\, m/\hbar$. The ergoregion is located where the isocontours are dashed lines. One can note that $g_{00}$ never changes sign on the axis so that ergoregions have always the shape of a torus.

\subsection{Rotating models with self-interacting potentials}\label{s:nonfree}

Having explored the free field, we turn now to scalar fields with some self-interaction, 
i.e. with terms beyond the mass one in the potential $V(|\Phi|^2)$. 
A great variety of such potentials have been proposed in the literature (see the reviews 
\cite{SchunM03,LieblP12}). An exhaustive study of them is beyond the scope of this paper.
We focus instead on two potentials mentioned in Sec.~\ref{s:intro}: the $\Lambda |\Phi|^4$ one
[Eq.~(\ref{e:V_phi_4})] and the solitonic one [Eq.~(\ref{e:V_sol})]. 

At lowest order the potentials (\ref{e:V_phi_4}) and (\ref{e:V_sol}) reduce to the free field one. The technique used to compute solutions follows from this property. One starts with one of the free-field solution and one slowly changes the potential to reach the desired value of the parameters $\Lambda$ [potential (\ref{e:V_phi_4})] or $\sigma$
[potential (\ref{e:V_sol})]. This technique works better if the starting configuration corresponds to small values of the scalar field, i.e. for values of $\omega$ close to $m/\hbar$. The precision of the obtained configurations can be assessed by the error indicators presented in Sec.~\ref{ss:errors}.

\begin{figure}
\includegraphics[width=8cm]{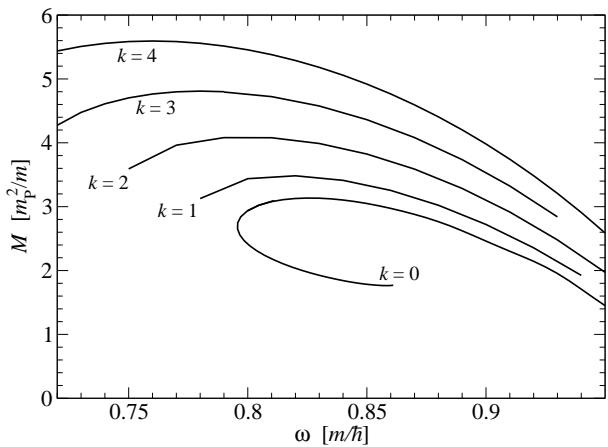}
\caption{ \label{fig:adm_lambda} Gravitational mass as a function of $\omega$
for boson star models constructed upon   
the self-interacting potential (\ref{e:V_phi_4}) with $\Lambda=200$, 
with different values of the rotational quantum number $k$.}
\end{figure}

Let us first consider the $\Lambda |\Phi|^4$ potential (\ref{e:V_phi_4}). 
In geometrized units, the constant $\Lambda$ is dimensionless. We select $\Lambda=200$, which 
is a representative value considered by Colpi et al. for their study of the nonrotating case
\cite{ColpiSW86}. The value of $\Lambda$ is chosen only for illustrative purposes and does not come from any physical motivation. In particular
it does not lead to a very big increase of the maximum mass.
 We have computed sequences of rotating configurations for this model, for
$k$ ranging from 1 to 4 ($k=0$ is also shown). 
Figure~\ref{fig:adm_lambda} shows the resulting gravitational mass $M$ as a function of $\omega$.
As a check, for $k=0$, we recover the maximum mass found by Colpi et al. \cite{ColpiSW86} for 
$\Lambda=200$: $M_{\rm max}^{(k=0)} = 3.14 \, m_{\rm P}^2/m$. 
For $k\geq 1$, we find the values given in Table~\ref{t:phi4_Mmax}.

\begin{table}
\begin{tabular}{l|ccccc}
\hline
$k$ & 0 & 1 & 2 & 3 & 4 \\
\hline
$M_{\rm max}$ [$m_{\rm P}^2/m$] & 3.14 & 3.48 & 4.08 & 4.81 & 5.59 \\
$\omega$ [$m/\hbar$] & 0.83 & 0.82 & 0.80 & 0.78 & 0.76 \\
\hline
\end{tabular}  
\caption{ \label{t:phi4_Mmax}
Maximum mass of rotating boson star models 
constructed upon the potential (\ref{e:V_phi_4}) with $\Lambda=200$.
The second line gives the value of $\omega$ for which the maximum
mass is reached.}
\end{table}

\begin{figure}
\includegraphics[width=8cm]{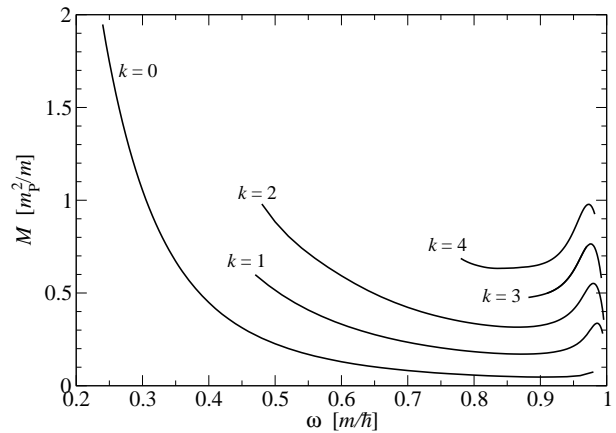}
\caption{ \label{fig:adm_eta} 
Gravitational mass as a function of $\omega$ 
for boson star models constructed upon the self-interacting potential 
(\ref{e:V_sol}) with $\sigma = 0.05$, with
different values of the rotational quantum number $k$.}
\end{figure}

For the solitonic potential (\ref{e:V_sol}), we perform the study for $\sigma=0.05$. As for $\Lambda |\Phi|^4$ potential this is
only an illustrative value.
Figure~\ref{fig:adm_eta} shows the results regarding the gravitational mass $M$. 
For solitonic boson stars, most of the previous works are concerned with relatively small values of $\omega$, where a maximum mass is observed. For these configurations the scalar field 
 $\phi$ is very close to a step function. Such a behavior is rather difficult to describe with spectral methods, for which smooth fields are required. Our code has therefore some difficulties in 
reaching very small values of $\omega$. For moderate values of $\omega$, our results are in good agreement with previous works. 
In particular, a secondary maximum is observed for $\omega$ near $m/\hbar$, as in 
Ref.~\cite{KleihKL05}.

This part of the article is not intended to constitute a detailed study of interacting potentials. It must rather be viewed as an illustration of the fact that our code is flexible enough to cope with various different situations. Comprehensive studies of the parameter space would however require some tuning of the various computational parameters. Let us eventually mention that no ergoregions 
have been found in the configurations presented in Figs.~\ref{fig:adm_lambda} and \ref{fig:adm_eta}.


\section{Timelike Geodesics}\label{s:orbits}

A standard way to analyze a given spacetime geometry is to study its geodesics. 
Moreover computing geodesics leads to astrophysical observable.  
Being interested in the motion
of stars in the vicinity of a supermassive boson star, 
we consider only timelike geodesics in this article, 
i.e. orbits of test particles of mass $\mu>0$.
In addition,  we restrict ourselves
to the equatorial plane ($\theta=\pi/2$) for simplicity.
In this Section we consider only free-field boson stars, i.e. the 
models computed in Secs.~\ref{s:results_spher}, \ref{s:resu_free_field}
and \ref{s:ergo}. 

The stationarity and axisymmetry of the underlying spacetime imply
the existence of two constants of motion along any geodesic. Given 
the components $p_\alpha$ of the particle's 4-momentum with respect to the
coordinates $(t,r,\theta,\varphi)$, these constants are expressible as 
$E = -p_{t}$ and $L=p_{\varphi}$ and are called respectively
the particle's energy ``at infinity'' and its angular momentum ``at infinity''. 
The equation governing the variation of the radial coordinate $r$ along 
an orbit in the equatorial plane can be written as (see e.g. Ref.~\cite{Gourg10})
\be
\label{e:veff}
\left(\frac{\D r}{\D\tau}\right)^{2}=\mathcal{V}\left(r,\varepsilon,\ell\right) ,
\ee
where $\tau$ is the particle's proper time and the effective potential in the radial direction, $\mathcal{V}$, 
is given by
\be
\mathcal{V}\left(r,\varepsilon,\ell\right)=\frac{1}{A^{2}}\left[\frac{1}{N^{2}}\left(\varepsilon+\beta^\varphi\ell\right)^{2}-\frac{\ell^{2}}{B^{2}r^{2}}-1\right],
\ee
with $\varepsilon:=E/\mu$ and $\ell:=L/\mu$. Given (\ref{e:veff}), $\mathcal{V}$ must be positive, 
which occurs if, and only if,
\be \label{e:epsilon_range}
    \varepsilon\leq\varepsilon_{\rm neg}
    \quad\mbox{or}\quad
    \varepsilon\geq\varepsilon_{\rm min},
\ee
where $\varepsilon_{\rm neg}$ and $\varepsilon_{\rm min}$ are the two roots 
of the equation\be
\label{e:veff0}
\mathcal{V}=0\iff\frac{\varepsilon^{2}}{N^{2}}+\frac{2\ell \beta^\varphi}{N^{2}}\varepsilon+\left(\frac{(\beta^\varphi)^2}{N^{2}}-\frac{1}{B^{2}r^{2}}\right)\ell^{2}-1=0
\ee
For any given value of $\ell$, this second order polynomial 
equation in $\varepsilon$ has for discriminant 
\be \label{e:discriminant}
\Delta=\left(\frac{2\ell \beta^\varphi}{N^{2}}\right)^{2}+ \frac{4}{N^{2}}\left\{
1 +  \left[ \frac{1}{B^{2}r^{2}} - \frac{(\beta^\varphi)^2}{N^{2}} \right]\ell^{2}\right\} ,
\ee
which is always positive. Thus (\ref{e:veff0}) does admit two solutions: 
\be
\label{e:emin}
\varepsilon_{\rm neg}=-\ell \beta^\varphi-\frac{N^{2}}{2}\sqrt{\Delta}
\quad\mbox{and}\quad
\varepsilon_{\rm min}= -\ell \beta^\varphi+\frac{N^{2}}{2}\sqrt{\Delta} .
\ee
Here we may distinguish two cases.
First of all, for boson stars without any ergoregion, 
the term in square brackets in Eq.~(\ref{e:discriminant}), 
is always positive (compare to Eq.~(\ref{e:ergo_criterion})), 
so that $\sqrt{\Delta} > 2\ell|\beta^\varphi|/N^2$ and 
$\varepsilon_{\rm neg}$ is always negative. 
Since in the absence of ergoregion, one has always $\varepsilon>0$, 
we conclude that in this case, only
the second inequality holds in (\ref{e:epsilon_range}). 
But, as discussed in Sec.~\ref{s:ergo}, 
very relativistic rotating boson stars may have ergoregions.
All signs are allowed for $\varepsilon$ in these regions, so in this
case we have to consider the two inequalities in (\ref{e:epsilon_range}). 
We treat these two cases separately in the next two subsections. 

\begin{figure}[!hbtp]
\includegraphics[width=8cm]{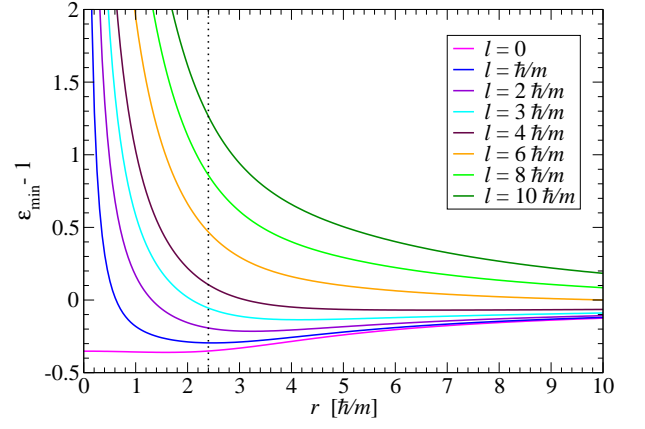}
\caption{ \label{fig:Emin}
Effective potential profiles for 
a free-field rotating boson star with $\omega=0.8\: m/\hbar$ and $k=1$. 
The vertical line represents the location of maximum of the scalar field
modulus $\phi$ (cf. Fig.~\ref{fig:contours}).
}
\end{figure}

\subsection{Effective potential outside ergoregions}

\begin{figure}[!hbtp]
\includegraphics[width=7cm]{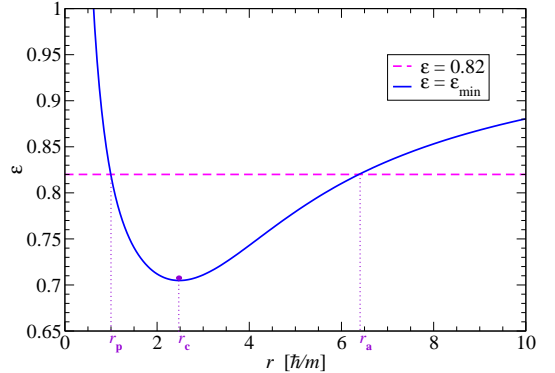}
\caption{ \label{fig:EminInter}
Effective potential profile for $\ell=1\:\hbar/m$. 
Only the region where $\varepsilon\geq\varepsilon_{\rm min}$ 
is allowed for the motion of the test particle, so the radial coordinate $r$ 
must obey $r_{\rm p}\leq r\leq r_{\rm a}$. The spot marks the position of the stable circular orbit of radius $r_{\rm c}$.
}
\end{figure}

To have a better understanding of the effective potential, we plot
$\varepsilon_{\rm min}$ as a function of $r$ for different values of $\ell$ 
in Fig.~\ref{fig:Emin}. Each extremum of these curves
corresponds to a circular orbit. 
We illustrate this fact in Fig.~\ref{fig:EminInter}, which
is the reproduction of the Fig.~\ref{fig:Emin} for a single value of $\ell$
($\ell=\hbar/m$). Indeed, we know that $\varepsilon$ is constant 
along the geodesic, so we choose an arbitrary value of $\varepsilon$
 and represent it by an horizontal dotted line in Fig.~\ref{fig:EminInter}.
The interval where $\varepsilon\geq\varepsilon_{\rm min}$ gives the allowed
values of the radial coordinate $r$ (in Fig.~\ref{fig:EminInter}, $r_{\rm p}\leq r\leq r_{\rm a}$)
and the values of $r$ for which $\varepsilon=\varepsilon_{\rm min}$ are turning
points, corresponding to the periastron and the apoastron.  If we choose the energy of the particle to be equal to the
minimum of $\varepsilon_{\rm min}\left(r\right)$, only one value of the
radial coordinate is allowed (in Fig.~\ref{fig:EminInter} it corresponds to $r_{\rm c}$):
this is a circular orbit. \\

We can infer two major facts from Fig.~\ref{fig:Emin}: first, all circular
orbits around mini boson stars are stable, since they correspond
to a minimum of the effective potential. 
Next, if we look at the
behavior of particles approaching the boson star, 
we see that for
$\ell\neq0$, there is a infinitely high potential wall 
preventing the particle to reach $r=0$.
But for the particles
with zero angular momentum ($\ell=0$), we have a finite value for
the energy at the exact center of the boson star. 
Note that we are assuming no interaction between the particle and the scalar 
field but the gravitational one, so that the particle may freely penetrate ``inside''
the boson star and reach its center. 
 
\subsection{Effective potential in ergoregions}

As discussed in Sec.~\ref{s:ergo}, 
rotating boson stars with ergoregions are unstable. If the (unknown) 
instability time scale in larger than the age of the Universe, then it is 
astrophysically relevant to study orbits around such stars, and in particular
inside the ergoregion. 

\begin{figure}[!hbtp]
\includegraphics[width=8cm]{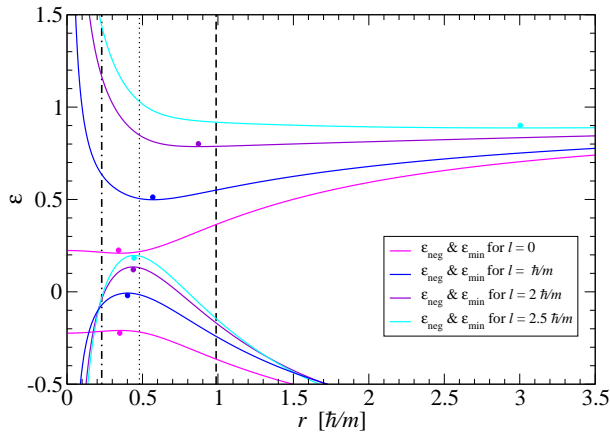}
\caption{ \label{fig:Eabos09}
Effective potential profiles for the free-field rotating boson star with $\omega=0.646\: m/\hbar$ and $k=1$. The ergoregion is delimited by the vertical dot-dashed and dashed lines, the thin vertical dotted line marks the maximum of the scalar field modulus $\phi$. 
The plots for $\varepsilon \geq 0.2$ are those of $\varepsilon_{\rm min}$, while 
the plots for $\varepsilon \leq 0.2$ are those of $\varepsilon_{\rm neg}$. The latter ones
can be used to determine orbits only inside the ergoregion. There is a circular stable orbit
(marked with a dot) for each minimum of $\varepsilon_{\rm min}$ and for each maximum of $\varepsilon_{\rm neg}$.
}
\end{figure}

In the ergoregion, $\varepsilon = E/\mu =-p_t/\mu = - p_\nu \xi^\nu / \mu$ can
be negative because in this part of spacetime the Killing vector associated
with stationarity $\xi = \partial / \partial t$
becomes spacelike. This is why in the ergoregion, we have to consider the two
solutions (\ref{e:emin}). Accordingly in Fig.~\ref{fig:Eabos09} we plot both 
$\varepsilon_{\rm min}\left(r\right)$
and $\varepsilon_{\rm neg}\left(r\right)$, for different values of $\ell$
and for a boson star that possesses an ergoregion. To develop an ergoregion,
the boson star spacetime must be very relativistic and such configurations
are obtained for small $\omega$. For Fig.~\ref{fig:Eabos09},
we chose $\omega=0.646\, m/\hbar$, along with
the rotational quantum number $k=1$. 
By ``inverting'' the reasoning made on Fig.~\ref{fig:EminInter}, taking into
account that $\varepsilon < \varepsilon_{\rm neg}$, we may say 
that the maximum of $\varepsilon_{\rm neg}$ corresponds to a stable circular 
orbits. These orbits, which exist only inside the ergoregion, are denoted 
by a dot in Fig.~\ref{fig:Eabos09}. We note also that inside the ergoregion, 
$\varepsilon_{\rm neg}$ becomes positive in some range of $r$ for $\ell$ 
large enough. 

\subsection{Circular orbits}

\begin{figure}[!hbtp]
\includegraphics[width=8cm]{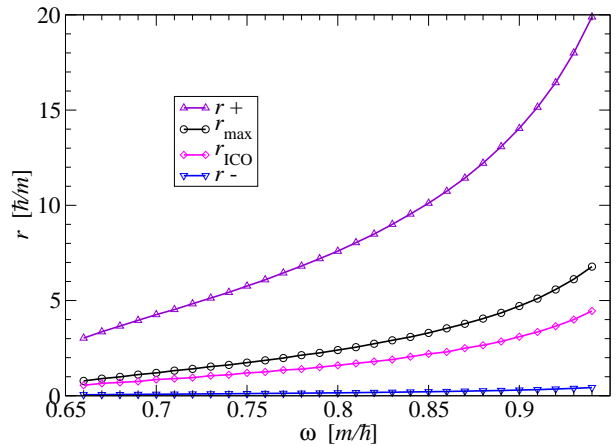}
\caption{ \label{fig:rmin}
Innermost circular orbit radius $r_{\textrm{ICO}}$  as a function of 
$\omega$ for boson stars with $k=1$. $r_{\textrm{max}}$
is the location of the maximum of the scalar field modulus $\phi$, while $r_{+}$ and $r_{-}$ 
denote to the radii where $\phi=\phi_{\textrm{max}}/10$.
}
\end{figure}

The two conditions satisfied by circular orbits are $\mathcal{V}=0$ and $\partial\mathcal{V}/\partial r=0$, these two equations admit two solutions written here in terms of the circular orbit velocity 
with respect to the \emph{zero angular momentum observer} or \emph{ZAMO}
(i.e. the observer of 4-velocity $n^\alpha$, cf. \cite{Bardeen70} and \cite{Gourg10}), 
which is given by 
\be \label{e:V_orbit_ZAMO}
V_{\pm}=\frac{\displaystyle -\frac{Br}{N}\displaystyle\frac{\partial \beta^\varphi}{\partial r}\pm\sqrt{D}}{2\left(\displaystyle \frac{1}{r} + \frac{1}{B}\frac{\partial B}{\partial r}\right)}
\ee
with 
\be
D:=\frac{B^{2}r^{2}}{N^{2}}\left(\frac{\partial \beta^\varphi}{\partial r}\right)^{2}+4\frac{\partial\nu}{\partial r}\left(\frac{1}{B}\frac{\partial B}{\partial r}+\frac{1}{r}\right) . 
\ee
$V_{+}$ is the velocity of the direct orbit and $V_{-}$ the velocity
of the retrograde one. For these solutions to exist we must have $D\geq0$.
We solved numerically this inequality for many boson stars and
found that there is a minimum radius under which no circular orbit can exist. 
Let us call the corresponding orbit the \emph{innermost circular orbit (ICO)}. 
The value $r_{\textrm{ICO}}$ of its $r$-coordinate depends on
the boson star as shown in Fig.~\ref{fig:rmin}. Let us point out that the ICO
is always located inside the torus.\\

Since $V_+$ and $V_-$ are velocities measured by a physical observer, the ZAMO, 
they must be subluminal, i.e. obey $\left|V_{\pm}\right|<1$. 
This criterion is always verified for boson stars.
 
\begin{figure}[!hbtp]
\includegraphics[width=8cm]{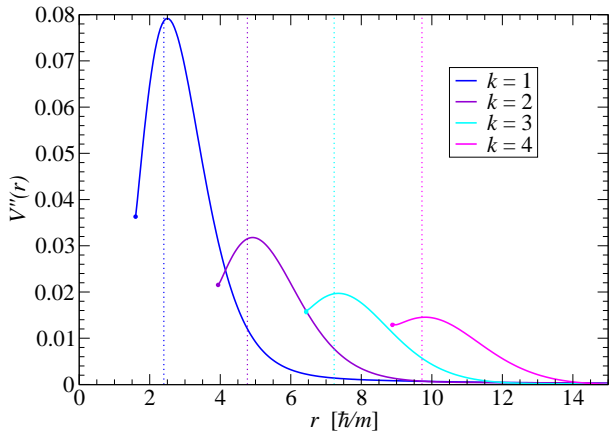}
\caption{ \label{fig:Stabcirc}
$\mathcal{V}"\left(r\right)$ for rotating boson stars with $\omega=0.8\: m/\hbar$. We choose $\varepsilon=4.2$ and $\ell=1.2\,\hbar/m$. The vertical lines mark the position of 
the maximum of the scalar field modulus for each boson star.
}
\end{figure}

Let us discuss now in more details the stability of the circular orbits. 
Circular orbits are stable if, and only if,
\be
\frac{\partial^{2}\mathcal{V}}{\partial r^{2}}=\mathcal{V}"\left(r\right)>0
\ee
If we plot $\mathcal{V}"\left(r\right)$ for various boson stars,
as we do for four of them with fixed values of $\varepsilon$ and $\ell$
in Fig.~\ref{fig:Stabcirc}, we see that this inequality is always strictly verified.
Thus, it seems that for rotating free-field boson stars, as long as a circular
orbit exists, it is stable. \\

We can compare these results to the black hole case (conditions for
existence and stability of circular orbits are globally the same for
Schwarzschild and Kerr spacetimes). For black hole spacetimes, circular orbits exist 
for $r$ larger than a critical value, so there is an ICO, 
located at $r_{\textrm{ICO}}=\left(1+\sqrt{3}/2\right)\,M\simeq1.87\, M$
for Schwarzschild spacetime. But the non-existence of circular orbits in certain regions of spacetime has a different cause for boson stars and black holes: for boson star this is due 
to $D$ becoming negative in the formula defining $V_{\pm}$ [Eq.~(\ref{e:V_orbit_ZAMO})]. 
On the contrary, $D$ is always positive outside the horizon of a black hole. 
In this case, it is $\left|V_{\pm}\right|$ that is becoming larger than one beyond the ICO
and thus prevent the existence of physical orbits. Moreover, the
 circular orbits around black holes, contrary
to those around boson stars, are stable only for $r\geq r_{\textrm{ISCO}}$
where ISCO stands for \emph{innermost stable circular orbit} and corresponds
to $r_{\textrm{ISCO}}=\left(5/2+\sqrt{6}\right)\,M\simeq4.95\,M$
in Schwarzschild spacetime\footnote{Let us recall that we are using
isotropic coordinates, not areal ones; for the latter Schwarzschild ISCO 
is located at the well-known value 
${\tilde r}_{\textrm{ISCO}}=6\,M$.}. For boson stars, we have found that, as 
long as a circular orbit exists, it is stable.

\subsection{Zero angular momentum orbits (\textmd{\normalsize $\ell=0$})}

\label{s:pointy}

\begin{figure}[!hbtp]
\includegraphics[width=8cm]{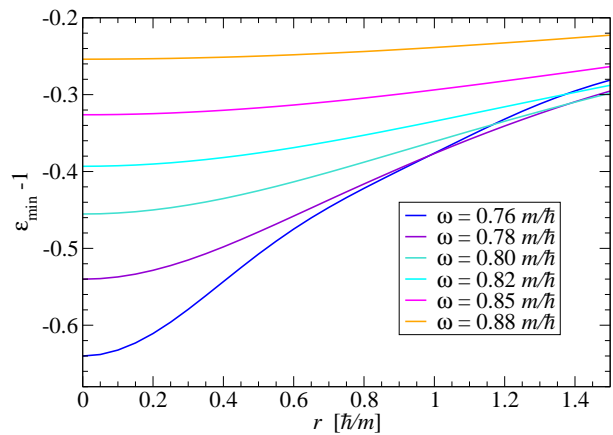}
\caption{ \label{fig:StatveffL0}
Effective potentials for $\ell=0$ and for spherical boson stars with different 
values of $\omega$ (decreasing from top to bottom).}
\end{figure}

\begin{figure}[!hbtp]
\includegraphics[width=8cm]{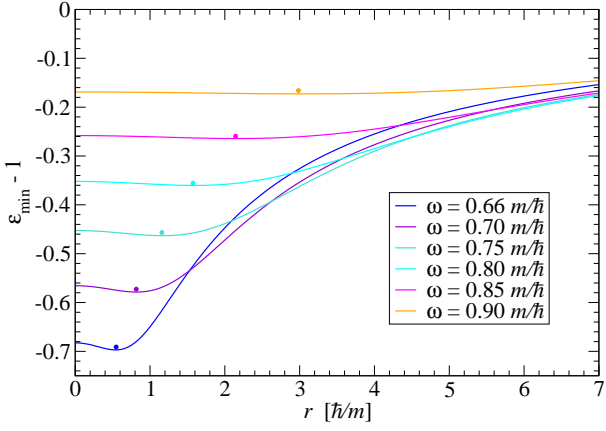}
\caption{ \label{fig:veffk1L0}
Effective potential for $\ell=0$ and for rotating boson stars with $k=1$ and different 
values of $\omega$ (decreasing from top to bottom). The dots mark the position of the stable circular orbits.
}
\end{figure}

\begin{figure}[!hbtp]
\includegraphics[width=8cm]{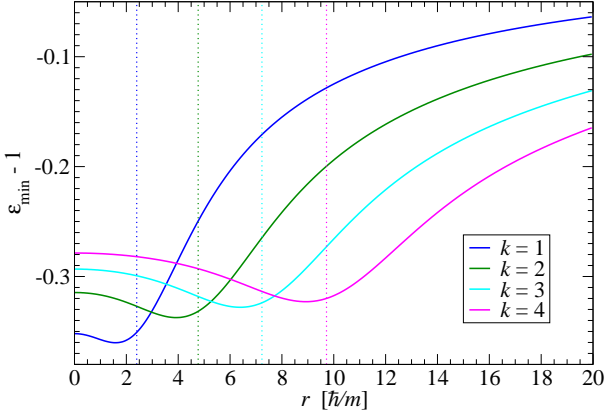}
\caption{ \label{fig:veffkL0}
Effective potential for $\ell=0$ and for rotating boson stars with $\omega=0.8\: m/\hbar$ and different values of $k$. The vertical lines mark the position of 
the maximum of the scalar field modulus for each boson star.
}
\end{figure}

The orbits with zero angular momentum are interesting in the case
of boson star because the particle is allowed to go through the star. This
is not the case for black holes or ordinary stars, where a particle
approaching towards the compact object faces either
the event horizon or the stellar surface. Let us express the effective
potential in the specific case $\ell=0$. Equation~(\ref{e:emin})
shows that $\varepsilon_{\rm min}$ is then equal to the lapse function:
\be
\varepsilon_{\rm min}=N \qquad (\ell=0).
\ee
We have plotted the profile of $\varepsilon_{\rm min}(r)$ in Fig.~\ref{fig:StatveffL0}
for different spherical ($k=0$) free-field boson stars. As noticed earlier, there is always
a stable equilibrium position at the center of the boson star. Then we consider rotating
boson stars, and in Fig.~\ref{fig:veffk1L0} we plot the effective potential for different
boson stars with a same value of $k$ ($k=1$) to compare with the
nonrotating case. We see that the equilibrium position still exists at
the center of the torus but has become unstable due to rotation. We
also note the existence of stable circular orbits close to the center
and remaining ``inside'' the boson star. In Fig.~\ref{fig:veffkL0}, we plot the
effective potential for boson stars with the same value of $\omega$
but for different values of $k$. The global behavior is the same
as in the previous figures but the scale enlarges as $k$ is increased. 
This is consistent with the increase of the size of the torus with $k$.\\

\begin{figure}[!hbtp]
\includegraphics[width=6cm]{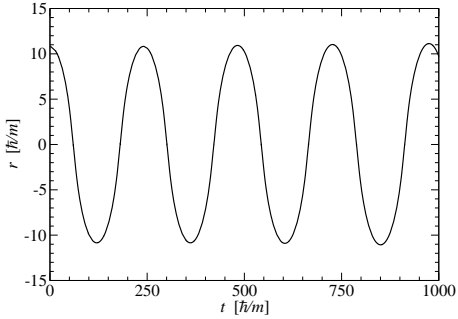}
\caption{ \label{fig:stat_bos_08}
Evolution of the $r$-coordinate of a $\ell=0$ test particle of initially at rest at 
$r_{i}=10.79\:\hbar/m$ in the spacetime generated by a spherical ($k=0$)
boson star with $\omega=0.77\: m/\hbar$.
}
\end{figure}

To fully determine the $\ell=0$ class of
orbits, we used the \texttt{GYOTO} code \cite{Vincent11,GYOTO,Vincent12} to integrate directly the geodesic equations within the 3+1 formalism \cite{Vincent12}, taking advantage of the capability 
of \texttt{GYOTO} to perform such an integration for a numerical metric. We first computed the
the geodesic of a particle initially at rest in the spacetime of a nonrotating spherically symmetric boson star ($k=0$). Due to spherical symmetry, the particle trajectory is a straight line. As expected, the
particle go straight through the center of the boson star and oscillate
back and forth, as it can be seen from Fig.~\ref{fig:stat_bos_08}, where 
we have plotted the particle's $r$-coordinate as a function of $t$.

\begin{figure}[!hbtp]
\includegraphics[width=8cm]{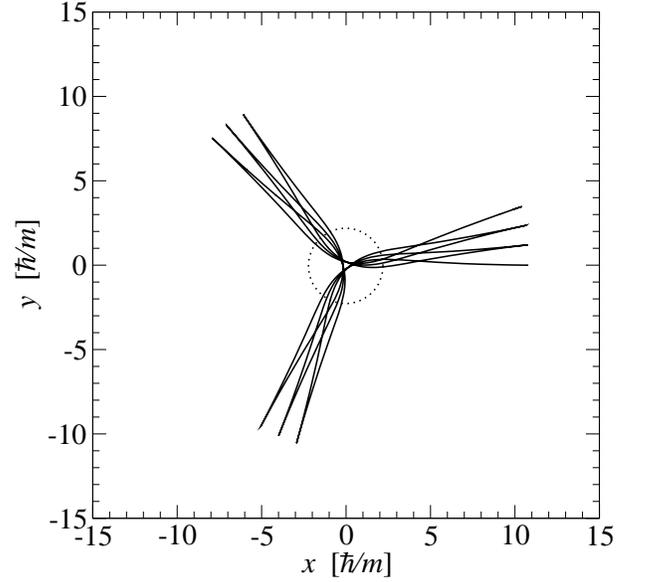}
\caption{ \label{fig:bos_1_08_phi0}
Orbit of a $\ell=0$ test particle in the equatorial plane of a rotating 
free-field boson star with $k=1$ and  $\omega=0.80\: m/\hbar$; the 
particle is initially at rest at $r = r_{\textrm{i}}=10.79\:\hbar/m$ and $\varphi=0$. 
The Cartesian-like coordinates of the plot are $x:=r\cos\varphi$ and 
$y:=r\sin\varphi$. The dotted circle marks the maximum
of the scalar field modulus $\phi$. 
}
\end{figure}

\begin{figure}[!hbtp]
\includegraphics[width=8cm]{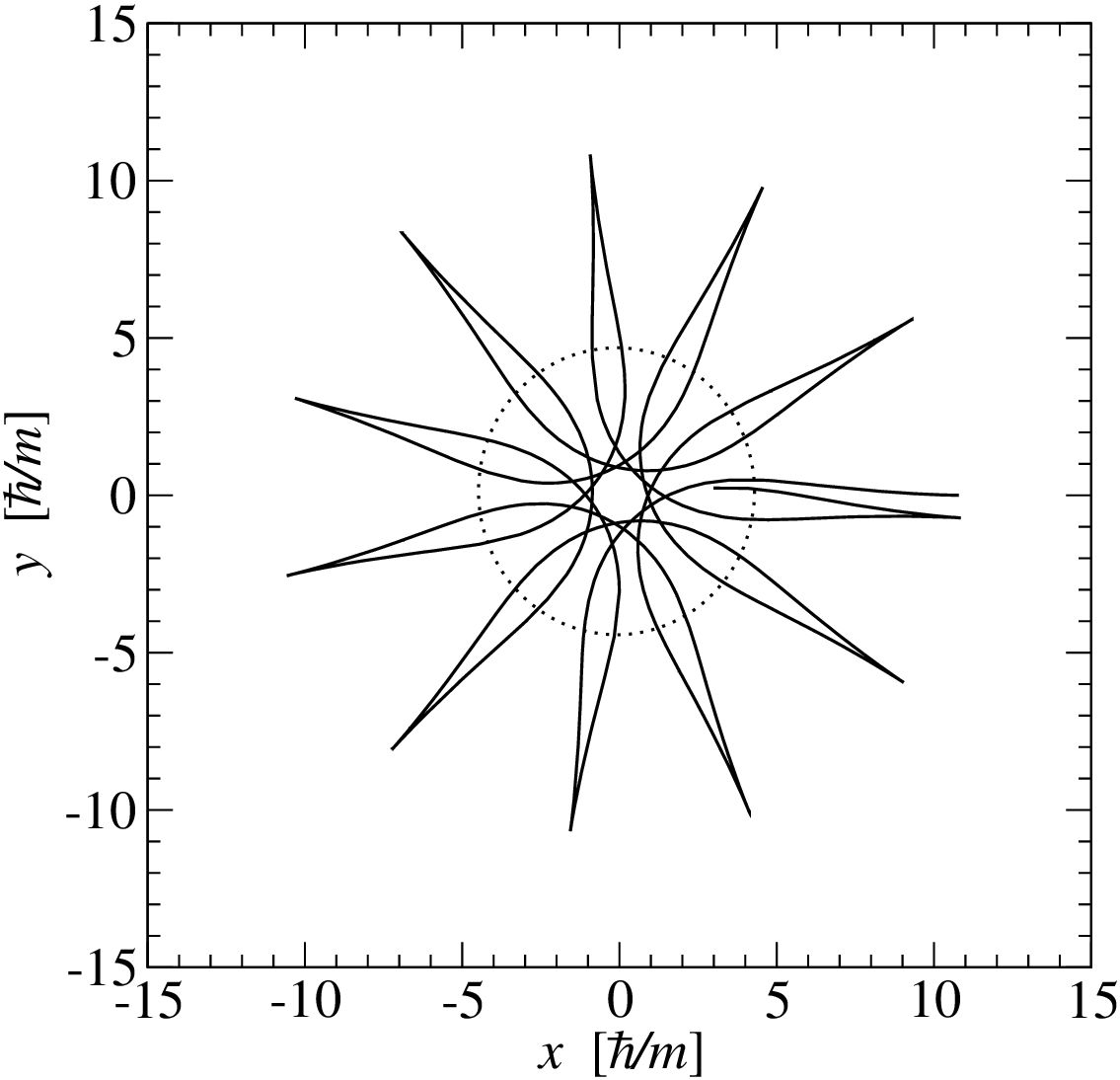}
\caption{ \label{fig:bos_2_08_phi0}
Same as Fig.~\ref{fig:bos_1_08_phi0} but for $k=2$. 
}
\end{figure}

We repeated this calculation for a rotating boson star with $k=1$ and $\omega=0.8\, m/\hbar$,
still starting at rest from $r_{i}=10.79\:\hbar/m$. The result is shown in 
Fig.~\ref{fig:bos_1_08_phi0}. We see clearly the manifestation of the Lense-Thirring
effect: the particle radially infalling is deflected near the center
and continues in almost a straight line before going backwards and
falling towards the center again. This gives rise to the
spike-like structure of the trajectory. Besides, this orbit is not closed.
In order to understand how these orbits are modified as the star's rotational 
quantum number $k$ is increased,
we plot the zero-angular-momentum orbit around boson stars with the
same value of $\omega$ as in Fig.~\ref{fig:bos_1_08_phi0}
but with $k=2$ in Fig.~\ref{fig:bos_2_08_phi0}
and  $k=3$ in Fig.~\ref{fig:bos_3_08_phi0}. We still see the characteristic
spikes, but the particle approaches less and less the center.
To investigate the behavior with respect to $\omega$, we plot in 
Fig.~\ref{fig:bos_2_075_phi0} a boson star with $k=2$ and $\omega=0.75\, m/\hbar$ 
to compare with Fig.~\ref{fig:bos_2_08_phi0}. We see that the effect of $\omega$ is to
change the value of the deviation angle when going through the
center.\\

\begin{figure}[!hbtp]
\includegraphics[width=8cm]{bos_3_08_phi0.eps}
\caption{ \label{fig:bos_3_08_phi0}
Same as Fig.~\ref{fig:bos_1_08_phi0} but for $k=3$. 
}
\end{figure}

\begin{figure}[!hbtp]
\includegraphics[width=8cm]{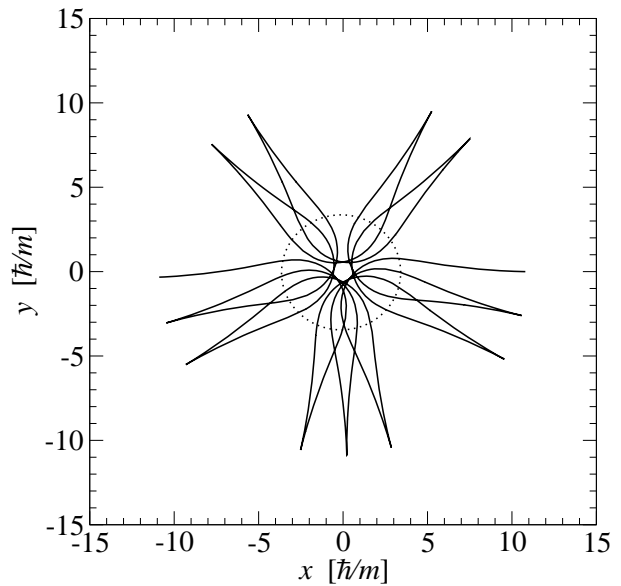}
\caption{ \label{fig:bos_2_075_phi0}
Same as Fig.~\ref{fig:bos_2_08_phi0} but for $\omega=0.75\: m/\hbar$.
}
\end{figure}

We may call the orbits displayed in Figs.~\ref{fig:bos_1_08_phi0}-\ref{fig:bos_2_075_phi0}
the \emph{pointy petal orbits}. Their spike-like structure 
is characteristic of boson star spacetimes, since
the particle has to be able to move very close to the center
to generate these orbits. This is indeed very different from 
the case of Kerr spacetime, in which all orbits are everywhere smooth
and where a particle initially at rest always 
falls into the black hole.

\section{Conclusions}\label{s:ccl}

We have developed a numerical code, based on multi-domain spectral methods, capable of 
solving the coupled Einstein-Klein-Gordon equations. We have used it to compute
models of rotating boson stars, with various self-interacting potentials 
for the scalar field.
We have obtained the first configurations with a rotational quantum number larger than 
$2$, namely $k=3$ and $k=4$. For $k=2$, we have determined the maximum 
mass of a free-field boson star: $M_{\rm max}^{(k=2)} = 2.216\,  m_{\rm P}^2/m$,
which was not known before. We have also confirmed the $k=1$ maximum mass found
by Yoshida and Eriguchi \cite{YoshiE97b}.  
For the self-interacting $\Lambda |\Phi|^4$ potential originally proposed
by Colpi et al. \cite{ColpiSW86}, we have computed the first rotating models, 
with $k$ ranging from $1$ to $4$ and have determined the corresponding 
maximum masses (cf. Table~\ref{t:phi4_Mmax}).

We have also numerically computed timelike geodesics in rotating boson star
spacetimes. In this article, we focused on circular orbits and 
zero-angular-momentum orbits. In particular, we have shown that as long
as $k\geq 1$, there is an innermost circular orbit (ICO), i.e. a 
radius $r_{\rm ICO}$ such that for $r < r_{\rm ICO}$ no circular orbit exist. 
For $r>r_{\rm ICO}$, all the orbits are stable. 
We have found that \emph{circular} orbits with 
zero angular momentum exist around boson stars, contrary to Kerr black holes. 
Moreover, we have exhibited a peculiar type of zero-angular-momentum orbits:
the pointy petal ones. Such orbits do not exist in black hole spacetimes. 
Therefore observing them around some astrophysical system, such 
as the Galactic Center, would be a strong indication in favor of a rotating 
boson star for the central compact object. 

In a future article \cite{Some_al14}, we shall perform a more 
systematic study of orbits around rotating boson stars, stressing the
differences with black holes. In particular, it will be interesting to know if 
one can allow non-zero angular momentum and still observe ``pointy petal orbits''
exhibited in Sec. \ref{s:pointy}. Preliminary studies seem to indicate that small 
but non negligible deviations in the value of $\ell$ are allowed.


\acknowledgments
We warmly thank Silvano Bonazzola, Thibaut Paumard and Fr\'ed\'eric Vincent for 
useful discussions and advices. EG acknowledges the support from the ANR grant 12-BS01-012-01
\emph{Analyse asymptotique en relativit\'e g\'en\'erale}.



\appendix

\section{Energy-momentum tensor} \label{s:app_T}

Let us derive the explicit expression of the scalar field energy-momentum 
tensor resulting from the ansatz (\ref{e:ansatzaxe}). 
First of all, (\ref{e:ansatzaxe}) yields to the following components with
respect to $(t,r,\theta,\varphi)$ coordinates:
\bea
    \nabla_\mu \Phi & = & (i\omega\phi,\;  \partial_r \phi,\;  \partial_\theta \phi,\; 
        -ik\phi) \exp[i(\omega t - k\varphi)]  \label{e:nab_Phi} \\
    \nabla_\mu \bar\Phi & = & (-i\omega\phi,\;  \partial_r \phi,\;  \partial_\theta \phi,\; 
        ik\phi) \exp[i(k\varphi-\omega t)] .   \label{e:nab_Phi_bar}
\eea
We may then evaluate $\nabla_\mu \Phi \nabla^\mu \bar{\Phi} = g^{\mu\nu} \nabla_\mu \Phi \nabla_\nu \bar{\Phi}$ by means of the 3+1 expression of $g^{\alpha\beta}$ (see e.g. Eq.~(5.51) of Ref.~\cite{Gourg12}), 
and get the explicit expression of the Lagrangian (\ref{e:scalar_Lag}):
\be
    {\mathcal L}_\Phi = \frac{1}{2} \left\{ \left[
    \frac{( \omega + k \beta^\varphi)^2}{N^2}  - k^2 \gamma^{\varphi\varphi} \right]  
    \phi^2 - g^{ab} \partial_a \phi \partial_b \phi - V \right\} ,  
\ee
where the indices $a$ and $b$ take the values $1$ and $2$ only (i.e. label the coordinates $(r,\theta)$). 
Plugging (\ref{e:nab_Phi})-(\ref{e:nab_Phi_bar}) into (\ref{e:tmunu}) leads to 
\bea
    T_{tt} & = & \omega^2 \phi^2 + {\mathcal L}_\Phi (-N^2 + \beta_\varphi \beta^\varphi) 
            \label{e:T_tt} \\
    T_{ta} & = & 0 \\
    T_{t\varphi} & = & - \omega k \phi^2 + {\mathcal L}_\Phi \beta_{\varphi}  \label{e:T_tp} \\
    T_{ab} & = & \partial_a \phi \partial_b \phi +  {\mathcal L}_\Phi  \gamma_{ab} \\
    T_{a\varphi} & = & 0 \\
    T_{\varphi\varphi} & = & k^2 \phi^2 + {\mathcal L}_\Phi  \gamma_{\phi\phi} , \label{e:T_pp}
\eea
where we have used the fact that $\beta_r=\beta_\theta=0$ and $\gamma_{a\varphi} = 0$
(circularity condition, cf. Sec.~\ref{s:rot_eq}). 

The trace of the energy-momentum tensor is obtained directly from (\ref{e:tmunu}):
\be \label{e:trace_T}
    T = g^{\mu\nu} T_{\mu\nu} = \nabla_\mu \Phi \nabla^\mu \bar{\Phi} + 4 
        \mathcal{L}_\Phi = 2 \mathcal{L}_\Phi - V . 
\ee

The 3+1 decomposition of the energy-momentum tensor lets appear the 
energy density $E$, the momentum density $P^i$ and the stress tensor
$S_{ij}$, the three of them as measured by the ZAMO
(i.e. the observer of 4-velocity $n^\alpha$). These quantities are obtained 
as the following projections of $T_{\alpha\beta}$:
\[
    E = T_{\mu\nu} n^\mu n^\nu ; \quad
    P_\alpha = - n^\mu T_{\mu\nu} \gamma^\nu_{\ \, \alpha}; \quad
    S_{\alpha\beta} = T_{\mu\nu} \gamma^\mu_{\ \, \alpha}
                        \gamma^\nu_{\ \, \beta} ,  
\]
with $\gamma^\alpha_{\ \, \beta} = \delta^\alpha_{\ \, \beta} + n^\alpha n_\beta$.
Given the components (\ref{e:T_tt})-(\ref{e:T_pp}) of $T_{\alpha\beta}$
and $n^\alpha = (1/N, 0, 0, - \beta^\varphi/N)$ and $n_\alpha = (-N,0,0,0)$,
we get
\be \label{e:E_axisym}
    E = \frac{1}{2} \left\{ \left[
    \frac{( \omega + k \beta^\varphi)^2}{N^2}  + k^2 \gamma^{\varphi\varphi} \right]  
    \phi^2 + \gamma^{ab} \partial_a \phi \partial_b \phi + V \right\} ,
\ee
\be \label{e:P_axisym}
    P_i = \left(0,\;  0,\; \frac{k}{N} (\omega+k\beta^\varphi) \phi^2 \right), 
\ee
\be \label{e:Sij_axisym}
    S_{ab} = \partial_a \phi \partial_b \phi +  {\mathcal L}_\Phi  \gamma_{ab} , \quad
    S_{a\varphi} =  0 , \quad
    S_{\varphi\varphi} =  k^2 \phi^2 + {\mathcal L}_\Phi  \gamma_{\phi\phi} .
\ee
The trace of the stress tensor, $S:= \gamma^{ij} S_{ij}$, has the following
expression:
\be \label{e:S_axisym}
    S = \frac{1}{2} \left\{ 3 \left[
    \frac{( \omega + k \beta^\varphi)^2}{N^2}  - k^2 \gamma^{\varphi\varphi} \right]  
    \phi^2 - \gamma^{ab} \partial_a \phi \partial_b \phi -3 V \right\} .
\ee


\end{document}